\documentstyle[12pt,epsf]{article}
\newcommand{\be}{\begin{equation}}
\newcommand{\ee}{\end{equation}}
\newcommand{\bea}{\begin{eqnarray}}
\newcommand{\eea}{\end{eqnarray}}
\newcommand{\bt}{\begin{tabular}}
\newcommand{\et}{\end{tabular}}
\newcommand{\beas}{\begin{eqnarray*}} 
\newcommand{\eeas}{\end{eqnarray*}}   
\newcommand{\ov}{\overline}

\newcommand{\bvec}{\mathbf}
\newcommand{\reac}{(\ref{reaction}) }
\def\ne{\hbox{$\nu_e \!$ }}
\def\neb{\hbox{$\ov{\nu}_e \!$ }}

\newcommand{\rt}{\rightarrow}

\newcommand{\gapproxeq}{\lower .7ex\hbox{$\;\stackrel{\textstyle >}{\sim}\;$}}
\newcommand{\lapproxeq}{\lower .7ex\hbox{$\;\stackrel{\textstyle <}{\sim}\;$}}

\def\z0{\mbox{$Z^0$}}

\def\ca{{C_{\scriptscriptstyle A}}}
\def\cv{{C_{\scriptscriptstyle V}}}

\def\caq{{C_{\scriptscriptstyle A}^2}}
\def\cvq{{C_{\scriptscriptstyle V}^2}}
\newcommand{\mod}{ \sum_{\mathrm spins} |M|^2 }

\newcommand{\tre}{ \left( \cvq + 3 \caq \right) }
\newcommand{\eqn}[1]{(\ref{#1})}
\newcommand{\wsig}{\widehat{\Sigma}}
\newcommand{\wsigp}{\widehat{\Sigma}'}

\newcommand{\slk}{/\!\!\!k}
\newcommand{\slp}{/\!\!\!p'}
\newcommand{\slq}{/\!\!\!q}

\begin{document}

\setlength{\unitlength}{1mm}

{\hfill astro-ph/9808196}

{\hfill DSF 30/98}\vspace*{2cm}

\begin{center}
{\Large \bf Precision Rates for Nucleon Weak Interactions in}
\end{center}
\begin{center}
{\Large \bf Primordial Nucleosynthesis and $ ^{4}He$ abundance}
\end{center}

\bigskip\bigskip

\begin{center}
{\bf S. Esposito}, {\bf G. Mangano}, {\bf G. Miele} and {\bf O. Pisanti},
\end{center}

\vspace{.5cm}

\begin{center}
{\it Dipartimento di Scienze Fisiche, Universit\'{a} di Napoli "Federico
II",
 and
INFN, Sezione di Napoli, Mostra D'Oltremare Pad. 20, I-80125 Napoli,
Italy}\\
\end{center}
\bigskip\bigskip\bigskip

\begin{abstract}
We report the results of a detailed calculation of nucleon weak
interactions relevant for the neutron to proton density ratio at the onset
of primordial nucleosynthesis. Radiative electromagnetic corrections,
finite nucleon mass terms, thermal radiative effects on weak processes and
on neutrino temperature are taken into account to reduce the theoretical
uncertainty on $n \leftrightarrow p $ rates to $1\%$. This translates into
a sensitivity in $^4 He$ mass fraction $Y_p$ prediction up to $10^{-4}$. We
find a positive total correction to the Born prediction $\delta Y_p
\simeq 0.004$.
\end{abstract}
\vspace*{2cm}

\begin{center}
{\it PACS number(s): 98.80.Cq; 95.30.Cq; 11.10.Wx; 13.40.Ks}
\end{center}

\newpage
\baselineskip=.8cm

\section{Introduction}
\setcounter{equation}0
Primordial Big Bang Nucleosynthesis (BBN) \cite{revBBN,Kolb}, represents
one of the greatest success of the hot Big Bang Model and is among the main
achievements of the interplay of particle physics and cosmology. The
predicted abundances of all light nuclei produced in the early stage of the
life of the universe, only depend on the baryon to photon density ratio
$\eta$, and a single value for $\eta \sim 10^{-10}$ may accomodate all
observed order of magnitude for $^{4}He$, $D$, $^{3}He$ and $^{7}Li$ data.
In the last few years, due to the increasing precision in observations
\cite{BBNdata}, an even more exciting possibility can be foreseen, namely
that BBN may tell us new pieces of information on both fundamental physics
beyond the Standard Model and our present understanding of the evolution of
the universe. In this respect it is already remarkable that BBN has been
able to strongly constraint the number of {\it effective} neutrino degrees
of freedom. This perspective, however, the fact that BBN is now entering in
his maturity and precision era, demands similar improvements in the
precision of the theoretical analysis, in order to reduce as much as
possible all uncertainties on the predictions. An increasing precision in
the measurements of $^{4}He$ mass fraction at the level of $10^{-4}$
\cite{HE4data} requires, for example, a reliability of proton-neutron
conversion rates at the level of one percent at least, and a control at the
same level of precision of all other effects which are relevant for the
neutron to proton ratio at the freeze-out.

BBN took place at temperatures in the range $0.1\div 10~ MeV$ and can be
seen as the final outcome of two stages, the decoupling of the weak
interactions which keep neutrons and protons in chemical equilibrium and,
shortly after, the onset of nuclear reactions which start producing light
nuclear species. At temperature $T >> 1~ MeV$ the neutron to proton ratio
is order unity and starts decreasing exponentially when the temperature
reaches the value of their mass difference. As the temperature decreases,
however, weak interactions are no longer fast enough to maintain
equilibrium and a substantial final neutron fraction survives down to the
phase of nucleosynthesis. All neutrons become practically bound in $^{4}He$
nuclei, due to the high binding energy per nucleon, which therefore
represent the dominant products of BBN, with an abundance in mass $Y_p$
close to 25 \%. While $Y_p$ is weakly depending on the baryon to photon
ratio $\eta$, it is instead strongly influenced by the ratio $n/p$ at the
freeze-out, so it is particularly relevant to have a good theoretical
estimate of this quantity.

The nuclear reaction chain starts with deuterium formation via $p+n
\rightarrow D + \gamma$ when the inverse rate for Deuterium dissociation
starts falling. This happens when the temperature has decreased below the
tiny D binding energy $E_B = 2.2 ~MeV$ and, due to the large photon to
baryon density ratio, when even the tail of photon Bose distribution
becomes ineffective in D photodissociation, i.e. $\eta^{-1} \exp \left(
-E_B/T
\right)\sim 1$, which gives $T \sim 0.1~MeV$. After Deuterium, the nuclear
chain starts, leading to sensible production of nuclei up to $^{7}Li$ (for
a recent review see \cite{Sarkar}).

The more recent data on $^{4}He$ mass fraction and D abundance are still
controversial since there are two different sets of results mutually
incompatible \cite{HE4data}
\begin{eqnarray}
Y_p & = &  0.234 {\pm} 0.0054 ~~~,~~~~~ D/H = (1.9 {\pm} 0.4 ) {\cdot} 10^{-4}~~~;
\nonumber
\\
Y_p &= & 0.243 {\pm} 0.003 ~~~,~~~~~ D/H = (3.40 {\pm} 0.25) {\cdot} 10^{-5}~~~;
\end{eqnarray}
which show the well known fact that large Helium mass fractions requires
low values for Deuterium abundance and viceversa. It is possible that
forthcoming measurements from high-redshift, low-metallicity QSO, or better
understanding of the present data will clarify the situation
\cite{BBNdata}, either in solving the controversy or pointing out a
difficulty of standard BBN scenario. The fact which however is emerging
from the above results is that Helium data are now reaching a precision of
one {\it per mille}, requiring a similar effort in reducing the theoretical
uncertainties on elementary $n \leftrightarrow p$ processes, as well as in
evaluating the effect of uncertainties in the experimental nuclear reaction
rates \cite{Nuclear}.

 The now {\it classical} numerical codes which are used to deduce nuclei
abundances \cite{code}, have been already modified by several authors, in
order to include corrections to the tree level $n \leftrightarrow p$
transition rates. These corrections have distinct nature and, listed as
they were historically introduced, are the following.

First of all, when accuracy of the order of one percent is required, all
Born amplitudes for $ n \leftrightarrow p$ processes should be corrected
for order $\alpha$ radiative corrections. These effects have been
extensively studied in literature \cite{Sirlin}-\cite{Wilk} and can be
classified in {\it outer} factors, involving the nucleon as a whole, and
{\it inner } ones, which instead depend on the details of nucleon internal
structure. Their inclusion in the $\beta$-decay rate reduces the tree level
estimate of neutron lifetime $\tau_n \simeq 961 ~sec$ to the value $\tau_n
\simeq 894~ sec$, which is closer to the experimental one
$\tau_n^{exp}= 886.7 {\pm} 1.9~sec$ \cite{PDG}. A similar correction, of the
order of 10 \%, is expected to affect the rates of all elementary processes
relevant for BBN
\begin{eqnarray}
(a)~~~ \nu_e + n \rightarrow e^- + p &~~~,~~~~~~& (d)~~~ \neb + p
\rightarrow e^+ + n~~~, \nonumber \\
(b)~~~e^- + p \rightarrow \nu_e + n &~~~,~~~~~~& (e) ~~~n \rightarrow e^- +
\neb + p ~~~,\nonumber \\
(c)~~~ e^+ + n \rightarrow \neb + p &~~~,~~~~~~& (f)~~~ e^- +
\neb + p \rightarrow n~~~.
\label{reaction}
\end{eqnarray}
Other small effects are actually expected at higher order, since the
estimated value of $\tau_n$ is compatible with the experimental value at
4-$\sigma$ level only. These additional contributions are usually taken
into account by eliminating the coupling in front of the reaction rates
$(a)$-$(f)$ in favour of $\tau_n^{exp}$.

It was quite soon realized \cite{Dicus,Cambier} that since all reactions
\reac
take place in a thermal bath of electron, positron, neutrinos,
antineutrinos and photons, finite density and temperature radiative effects
should be included as well. They may affect the Born rates for factors of
the order $\alpha T/m_e$, with $m_e$ the electron mass, which in the
temperature interval relevant for BBN may be as large as few percent.
Calculations of these effects, using the real time formalism of finite
temperature field theory \cite{Dolan}, has been done by several authors
\cite{Dicus,Cambier}, \cite{Donoghue85}-\cite{EMMP}. Though they agree on
the magnitude of the induced effect on $^4He$ abundance, there are however
striking differences in the sign of the corrections.

Thermal effects play also a role in modifying neutrino temperature after
decoupling. The difference between neutrino and photon temperatures
originates from the entropy release to photon in $e^+$-$e^-$ pair
annihilation which, roughly speaking, takes place at temperature of the
order of the electron mass $m_e$, shortly after neutrino decoupling. The
interaction with the surrounding medium gives an extra plasma mass term to
$e^+$-$e^-$, of the order of $\alpha T^2/m_e$, which slightly changes the
neutrino to photon temperature ratio. Actually it has been shown that there
is a residual small coupling of neutrinos at the electron-positron
annihilation phase, which introduces non equilibrium effects in the
neutrino energy distribution \cite{Dicus},\cite{Dodturn}. This last effect
introduces corrections to $^{4}He$ abundance of the order of $(1 \div 2) {\cdot}
10^{-4}$.

Finally all Born amplitudes for the processes in \reac should be also
corrected for nucleon finite mass effects. This has been first considered
in \cite{main}. They affect the allowed phase space as well as the weak
amplitudes, which should now include the contribution of nucleon weak
magnetism. Initial nucleons with finite mass will also have a thermal
distribution in the comoving frame, over which it is necessary to average
the transition probabilities.

In this paper we have considered in detail all the above effects, with the
only exception of non equilibrium distortion of neutrino distribution,
whose contribution to $^{4}He$ mass fraction is however very small. In
particular:
\begin{itemize}
\item[i)] all radiative corrections have been included at order $\alpha$;
\item[ii)] thermal radiative effects have been recalculated;
\item[iii)] all finite nucleon mass corrections at order $1/M_N$ are considered.
\end{itemize}
The motivation for such an analysis which, as we have stressed, is
certainly not new, is twofold. On one side for some of these effects
different results have been obtained by several authors. Furthermore the
precision expected for $^4He$ mass fraction measurements requires a careful
scrutiny of the net effect of all these one percent level corrections.

We have organized the paper as follows. In section 2 we report the
customary expressions for Born rates for the processes \reac. Section 3 is
devoted to radiative zero temperature corrections. In section 4 we discuss
finite nucleon mass corrections, while in section 5 we report all our
results for thermal radiative effects, evaluated in the real time
formalism. In section 6 we present our results and discuss the expected
correction to $^{4}He$ mass fraction. Finally in section 7 we report our
conclusions and outlook.

\section{The Born rates for $n \leftrightarrow p$ reactions}
\setcounter{equation}0

At the epoch of BBN, $0.1~MeV \leq T \leq 10~MeV$, the universe was filled
by a primordial plasma of nucleons, electrons, positrons, neutrinos,
antineutrinos and photons, which were initially in thermal equilibrium. The
initial conditions for the primordial nucleosynthesis are settled out by
the ratio of the relative abundances of neutrons and protons. For
temperature larger than $1~MeV$ this ratio is strongly fixed by
thermodynamical equilibrium, but as $T$ continues decreasing the necessary
conditions to sustain the equilibrium are no longer satisfied. In this
regime, to obtain the neutron-proton abundances, it is necessary to solve
the Boltzmann equation, where the rates for the charged-current weak
interactions transforming $n
\leftrightarrow p$ \reac are involved. Shortly after, when the temperature
drops down to $0.1 \div 0.3~MeV$, the nuclear reactions become strongly
active and transform neutrons mainly in $^{4}He$, and only in a small
fraction in the other light elements.

To obtain the rates per nucleon for the $n \leftrightarrow p$ processes,
let us consider a generic transition $i \rightarrow f$ in
\reac. The differential reaction rate can be written in the form
\begin{eqnarray}
d \Gamma (i \rightarrow f) \, & = & \, \sum_{\mathrm spins} |M(i
\rightarrow f)|^2 \prod_i \frac{d^3 {\bvec{p}}_i}{(2 \pi)^3 2 p_{0i}} \, F_i
\prod_f \frac{d^3 {\bvec{p}}_f}{(2 \pi)^3 2 p_{0f}} \, (1- F_f) \nonumber \\
& {\times} & (2 \pi)^4 \delta^{(4)} \left( \sum_i p_i - \sum_f p_f \right)~~~,
\label{2.1}
\end{eqnarray}
where $|M(i \rightarrow f)|^2$ is the squared matrix element, to be summed
over all spin degrees of freedom, the four-momentum of $j$-th particle is
$p_j
\, =
\, (p_{0j}, {\bvec{p}}_j)$, and $F_j$ denotes the phase space density of
the $j$-th particle.

Since the temperatures involved during the BBN are less than $10~MeV$, in
first approximation one can neglect in \eqn{2.1} the kinetic energy of
nucleons with respect to their mass. Moreover the Pauli blocking factor for
the final nucleon is completely negligible, since nucleons are strongly
nonrelativistic and $F_{N_f} \leq \exp(-M_{N_f}/T) \leq \exp(-100)$, where
$N_f$ denotes the final nucleon. In this limit, the differential reaction
rate per incident nucleon ($N_i$) becomes
\begin{eqnarray}
d \omega_B (i \rightarrow f) \, & = &  \, 2 \pi \sum_{\mathrm spins}
\frac{|M(i \rightarrow f)|^2}{8 M_{N_i} M_{N_f}} \prod_{i \neq N_i} \frac{d^3
{\bvec{p}}_i}{(2 \pi)^3 2 p_{0i}} \, F_i \prod_{f \neq N_f} \frac{d^3
{\bvec{p}}_f}{(2 \pi)^3 2 p_{0f}} \, (1-F_f) \nonumber \\ & {\times} & \delta
\left( M_{N_i} - M_{N_f} +
\sum_{i \neq N_i} p_{0i} - \sum_{f \neq N_f} p_{0f} \right)~~~,
\label{2.3}
\end{eqnarray}
Hereafter, we will denote with $\omega(i
\rightarrow f)$ the reaction rate per incident nucleon for the process $i
\rightarrow f$ and the index $B$ reminds that the rates are computed at tree
level {\it and} in the infinite mass limit for nucleons. We will refer to
this as the Born approximation.

Let us start considering as an example the reaction (a) in \reac. At tree
level, the corresponding $V-A$ amplitude for the diagram of Figure 1 is
\begin{equation}\label{cb10}
M(\ne + n \rightarrow  e^- + p) \, = \, \frac{G_F}{\sqrt{2}} \, \ov{u}_p(p)
\gamma_\mu (\cv - \ca \gamma_5)   u_n(q') ~   \ov{u}_e(p') \gamma^\mu (1 -
\gamma_5) u_\nu(q),
\end{equation}
where $G_F$ is the Fermi coupling constant and $\cv$, $\ca$ are the vector
and axial coupling of the nucleon. In Born approximation we have for
\eqn{cb10}
\begin{equation}\label{cb11}
\sum_{\mathrm spins} \frac{|M|^2}{8 M_{n} M_{p}} = 4  G_F^2 \left[ \tre
p'_0  q_0+ \left( \cvq- \caq \right) \bvec{p}' {\cdot} \bvec{q}   \right].
\end{equation}
Thus, by substituting \eqn{cb11} in \eqn{2.3} and after integrating we get
\begin{equation}\label{cb15}
 \omega_B ( \ne  +  n  \rightarrow  e^-  +  p ) \, = \,  \frac{G_F^2
\tre}{2 \pi^3} \, \int_0^\infty d {|\bvec{p}'| \,|\bvec{p}'|^2} \,  q_0^2
\, \Theta(q_0) \,F_\nu (q_0) \left[ 1 - F_e(p_0') \right],
\end{equation}
where the integration limits are imposed by the $\Theta$-function, $q_0
\geq 0$. In the present case we have $q_0 = - (M_n - M_p) + p_0' \equiv -
\Delta + p_0'$.
The Fermi statistical distributions for electrons and neutrinos in the {\it
comoving frame} are \footnote{We are assuming vanishing chemical potential
for neutrinos. See \cite{Sarkar} for a review on the effects induced by
neutrino asymmetry and constraints on chemical potentials.}
\be
F_e(p_0') = \left[e^{\beta |p_0'|} +1 \right]^{-1}~~~,~~~~~~~~~ F_\nu(q_0)
=
\left[e^{\beta_\nu |q_0|} +1 \right]^{-1}~~~, \label{cbb}
\ee
with $\beta=1/T$ and $\beta_\nu = 1/T_\nu$.

Due to crossing symmetry,  for the other five processes in \reac the form
of $\mod$ in (\ref{2.3}) remains unchanged, so the final result can be
obtained from (\ref{cb15}) by simply replacing  $q_0$, determined by the
energy conservation for each reaction, and the thermal factors, which
depend on the initial and final lepton states for each process. In Table 1,
using the same notation adopted for $\ne  +  n \rt
 e^-  +  p$, we report the values for $q_0$ and the thermal factors for
each reaction in \reac$\!\!$, while in Figure 2 we show the Born rates for
processes \reac versus photon temperature.\\ The electron neutrino
temperature behaviour after decoupling is shortly reviewed in Appendix A.
The result quoted there for the ratio $T_\nu / T$ will be used all through
our analysis. Non-equilibrium effects on neutrino distributions due to
residual coupling during the $e^+$-$e^-$ annihilation phase, and their
implications on $^4He$ abundance will be shortly discussed in section 6.

\begin{center}{Table 1.}\\
\begin{tabular}{|c|r|c|}
  \hline \hline
  Reactions & $q_0~~~~$ & Thermal factors \\ \hline \hline
  $\ne + n \, \rt \, e^- + p$ & $- \, \Delta \, + \, p_0'$ &
  $F_\nu(q_0) \; (1  -  F_e(p_0'))$ \\
  $e^- + p \, \rt \, \ne + n$ & $- \, \Delta \, + \, p_0'$ &
  $F_e(p_0') \; (1  -  F_\nu(q_0))$ \\
  $e^+ + n \, \rt \, \neb + p$ & $\Delta \, + \, p_0'$ &
  $F_e(p_0') \; (1  -  F_\nu(q_0))$ \\
  $\neb + p \, \rt \, e^+ + n$ & $\Delta \, + \, p_0'$ &
  $F_\nu(q_0) \; (1  -  F_e(p_0'))$ \\
  $n \, \rt \, e^- + \neb + p$ & $\Delta \, - \, p_0'$ &
  $(1  -  F_\nu(q_0)) \; (1  -  F_e(p_0'))$ \\
  $e^- + \neb + p \, \rt \, n$ & $\Delta \, - \, p_0'$ &
  $F_\nu(q_0) \; F_e(p_0')$ \\ \hline \hline
\end{tabular}
\end{center}

\section{Radiative electromagnetic corrections: the neutron lifetime}
\setcounter{equation}0

In order to measure the level of validity of Born approximation in the
evaluation of rates \reac, one can use the corresponding prediction for
neutron lifetime. By using \eqn{cb15} and Table 1, in the limit of
vanishing density one gets the simple expression
\begin{equation}\label{cb17}
\tau_n^{-1} \, = \,  \frac{G_F^2 \tre}{2 \pi^3} \, m_e^5 \,
\int_1^{\frac{\Delta}{m_e}} \,  d \epsilon \, \epsilon \, \left( \epsilon
\, - \, \frac{\Delta}{m_e}  \right)^2 \, \left( \epsilon^2 \, - \, 1
\right)^{\frac{1}{2}}~~~.
\end{equation}
Thus, inserting the value of $G_F$, very well known from the measurement of
muon decay \cite{PDG}, $\cv = V_{ud}=0.9751{\pm}0.0006$ and $\ca/\cv=1.2601 {\pm}
0.0025$, deduced from the study of the decay product angular distribution
in neutron decay \cite{PDG}, we obtain the value $\tau_n \simeq 961\, sec$
which has to be compared with the experimental result $\tau_n^{exp}=886.7
\, {\pm} \, 1.9 \, sec$ \cite{PDG}. As a consequence, a correction factor to
\eqn{cb17} of about $8 \%$ is needed. Note that similar corrections
are expected to affect the other reactions in
\reac$\!\!$ as well, since they are mediated by the same interaction
hamiltonian.

Let us consider order $\alpha$ QED corrections to neutron lifetime
($\alpha$ is the fine structure constant); these can be separated into {\it
outer} corrections, involving the nucleon as a whole, and {\it inner}
corrections, depending on nucleon structure. Obviously, the inner
corrections are sensible to the details of the strong interactions inside
the nucleon, while it can be shown \cite{Sirlin} that outer corrections, at
least up to terms of order $\alpha$, take a general form.

In terms of Feynman diagrams the corrections at order $\alpha$ to neutron
lifetime are obtained from the pure $V-A$ interaction \eqn{cb10} with the
exchange of a photon between the four fermions involved. Note that nucleons
interact electromagnetically through both their charge and magnetic moment,
even though the interactions involving magnetic moments are suppressed by
the inverse of the nucleon mass.

The outer correction to the differential decay rate can be written as the
multiplicative factor
\begin{equation}\label{cb18}
\left[1+ \frac{\alpha}{2 \pi} \, g(p_0',q_0)\right]~~~,
\end{equation}
where $g(p_0',q_0)$ is a function of electron and neutrino energy, whose
explicit expression can be found in eq. $(20b)$ of Ref. \cite{Sirlin}. This
function describes, at order $\alpha$, the deviations in the allowed
electron spectrum arising from the radiative corrections\footnote{It is
customary to substitute the function $g(p_0',q_0)$ in (\ref{cb18}) with its
mean value $\ov{g}$ obtained averaging over the allowed electron spectrum.
However, since we will consider not only neutron decay, but also the other
five reactions in
\reac$\!\!$, we will not use this approximation.}. It increases the decay
probability for neutron $\beta$-decay (and therefore decreases neutron
lifetime) by about 1.6 \%.

Inner corrections are much more difficult to handle since they strictly
depend on nucleon structure. In general, one can follow two different
approaches. One possibility is to directly consider radiative corrections
to the effective nucleon weak current  $\ov{u}_p \gamma_\mu (1 - (\ca/\cv)
\gamma_5) u_n$ \cite{Kubo}. Alternatively, one can study corrections
to quark currents $\ov{q} \gamma_\mu (1 - \gamma_5) q$ and then translate
the quark-based description into the hadronic one \cite{Marciano}. Here, we
adopt the second point of view, and report the results obtained by Marciano
and Sirlin \cite{Marciano}
\begin{equation}\label{cb19}
\frac{\alpha}{2 \pi} \left( 4 \, \ln \frac{M_Z}{M_p} \, + \, \ln
\frac{M_p}{M_A} \, + \, 2 C \, + A_g \, \right)~~~.
\end{equation}
The first term represents the dominant model-independent short-distance
contribution. The second and third terms are axial-current induced
contributions, where $M_A$ is a low energy cutoff applied to the
short-distance part of the $\gamma W$ box diagram and $2 C$ is the
remaining long-distance (low energy) correction. These terms depend on the
details of the strong interaction structure and are the main sources of
uncertainty on the radiative corrections. The allowed range for the cutoff
$M_A$ is $400 \div 1600 \, MeV$, while $C \, = \, 0.798 ~\ca ~(1+
\mu_p + \mu_n)$, $\mu_p + \mu_n \, \simeq \, -0.12$ being the nucleon
isoscalar anomalous magnetic moment. Then we have
\begin{equation}\label{cb20}
\frac{\alpha}{2 \pi} \left( \ln \frac{M_p}{M_A} \, + \, 2 C  \right) \,
\simeq \, 0.0012 \, {\pm} \, 0.0018 ~~~.
\end{equation}
The last term in (\ref{cb19}) is a perturbative QCD correction whose
calculation is rather reliable and gives $ A_g \, \simeq \, - \, 0.34$.

The largest correction comes from the first term in (\ref{cb19}) and it
seems appropriate to approximate the effects of higher orders by summing
all leading-logarithmic corrections of the type $\alpha^n
\ln^n(M_Z)$  via a renormalization group analysis. This has been done in
\cite{MS}; the corrected differential rate then acquires the multiplicative
factor
\begin{equation}\label{cb22}
{\cal G}(p_0',q_0) = \left[ 1 \, + \,  \frac{\alpha}{2 \pi} \left( \ln
\frac{M_p}{M_A} \, + \, 2 C  \right) \, + \, \frac{\alpha (M_p)}{2 \pi} \,
\left[ g(p_0',q_0) \, + \, A_g \right] \right] S(M_p , M_Z)~~~,
\end{equation}
where $\alpha (\mu)$ is the QED running coupling constant defined in the
$\ov{MS}$ scheme.  The short-distance enhancement factor $S(M_p,M_Z)$ is
given by \cite{MS}, \cite{Marciano} \footnote{The expression reported in
\cite{Marciano} has been updated to account for the measured top quark
mass.}
\begin{equation}\label{cb25}
S(M_p,M_Z) \, = \, \left( \frac{\alpha(m_c)}{\alpha(M_p)} \right)^
{\frac{3}{4}}  \left( \frac{\alpha(m_\tau)}{\alpha(m_c)} \right)^
{\frac{9}{16}}  \left( \frac{\alpha(m_b)}{\alpha(m_\tau)} \right)^
{\frac{9}{19}}  \left( \frac{\alpha(M_W)}{\alpha(m_b)} \right)^
{\frac{9}{20}}  \left( \frac{\alpha(M_Z)}{\alpha(M_W)} \right)^
{\frac{36}{17}}~~~.
\end{equation}
In the $\ov{MS}$ scheme the running coupling constant at different scales
($\alpha^{-1} (M_Z)$, $\alpha^{-1} (M_W)$, $\alpha^{-1} (m_b)$,
$\alpha^{-1} (m_\tau)$, $\alpha^{-1} (m_c)$, $\alpha^{-1} (M_p)$) is
(127.90, 127.94, 132.08, 133.27, 133.62, 133.90) leading to
 $S(M_p,M_Z) \,\simeq \, 1.0225$.

The radiative corrections just considered, which are described by the
factor in (\ref{cb22}), are not, however, the only ones to be taken into
account for the neutron decay. They contribute for about 4 \% to the
lifetime, against the required global correction of about $8 \%$. The
remaining leading corrections are usually viewed as corrections to the
phase space factor \cite{Wilk}, namely to the differential rate in
(\ref{cb17}), since they are electron and neutrino energy dependent.

Let us first examine the correction due to the distortion of the outgoing
electron wave function by the Coulomb field of the proton. This can be
calculated by solving the Dirac equation for an electron under the
influence of a spatially finite proton charge distribution of radius $R
\simeq 1 \, fm$; the wave function is then evaluated at the centre of the
proton. The correction factor is then given by \cite{Wilk}
\begin{equation}\label{cb27}
{\cal F}(p_0') \, {\cal L}(p_0') \, \simeq \, \left( 1 \, + \, \alpha \pi
\, \frac{p_0'}{{|\bvec{p}'|}} \right) \left[ 1 \, - \, \alpha \, R \, p_0'
\left( 1+ \frac{m_e^2}{2 {p_0'}^2} \right) \right]~~~.
\end{equation}
The first term is usually denoted as the Fermi function for the Coulomb
scattering.

Another small correction comes from the observation that neither the
electron wave function, evaluated at the centre of the proton through
${\cal F}(p_0') \, {\cal L}(p_0')$, nor the anti-neutrino one are constant
through the proton volume. Thus, the decay rate has to be computed by an
appropriate convolution of the electron, anti-neutrino and proton wave
functions through the proton volume. This leads to the following correction
factor \cite{Wilk}
\begin{equation}\label{cb28}
{\cal C}(p_0',q_0) \, \simeq \, 1 \, + \, c_0 \, + \, \frac{c_1}{p_0'} \, +
\, c_2 \, p_0' \, + \, c_3 \, {p_0'}^2~~~,
\end{equation}
with
\bea
c_0 & = & \frac{1}{5} \left[ R^2 \, m_e^2 \left( 1 \, - \frac{2}{3} B \, -
\, \frac{(p_0'+q_0 )^2}{m_e^2} \right) \, - \, \alpha \, B \, R \,
(p_0'+q_0 ) \right]~~~,\label{cb28a}\\ c_1 & = & \frac{2B}{15} \, R^2 \,
m_e^2 (p_0'+q_0 ) ~~~,\label{cb28b}\\ c_2 & = & \frac{R}{5} \, (3 \, - \,
B) \, \left[ \frac{2}{3} \, R \, (p_0'+q_0 )\, - \, \alpha \right]
~~~,\label{cb28c}\\ c_3 & = & \frac{2}{15} \, (3 \, - \, B) \, R^2
~~~,\label{cb28d}\\ B   & = & \frac{\cvq \, - \,
\caq}{\tre}~~~.\label{cb28e}
\eea
Finally, there are also tiny corrections related to the fact that the
outgoing proton is not at rest. An exhaustive discussion of finite nucleon
mass corrections will be given in the following. Here, we only point out
that since proton recoils, its Coulomb field is the one of a moving source.
The correction associated to this effect is contained in the factor
\cite{Wilk}
\begin{equation}\label{cb29}
{\cal Q} (p_0',q_0) \, \simeq \, 1 \, - \, \frac{\alpha \, \pi \, m_e^2}
{M_p \, {|\bvec{p}'|}} \, \left( 1 \, + \, B \, \frac{q_0}{3 p_0'}
\right)~~~.
\end{equation}
We eventually get for the neutron $\beta$-decay differential rate,
corrected at order $\alpha$
\be
d \omega(n\rightarrow e^- + \neb + p) = {\cal G}(p_0',q_0) \, {\cal
F}(p_0')\, {\cal L}(p_0')\, {\cal C}(p_0',q_0) \,  {\cal Q}(p_0',q_0)~
d\omega_B(n\rightarrow e^- + \neb + p), \label{cb31}
\ee
and thus the neutron lifetime reads
\bea
\tau_n^{-1} & = & \frac{G_F^2 \tre}{2 \pi^3}  \,  \int_{m_e}^{\Delta} d
p_0' \, p_0' \,{|\bvec{p}'|}\, \left( p_0' \, - \, \Delta \right)^2
\nonumber \\
& {\times} & {\cal G}(p_0',q_0) \, {\cal F}(p_0') \, {\cal L}(p_0') \,
{\cal C}(p_0',q_0) \, {\cal Q}(p_0',q_0)~~~. \label{cb30}
\eea

Concerning the relevance of the different corrections just considered, a
comment is in turn. For seek of completeness we have reported an exhaustive
description of all corrections at order $\alpha$, but as clearly appears
from the explicit expressions of (\ref{cb22})-(\ref{cb29}), the main
contributions come from \eqn{cb22} and ${\cal F}(p_0')$ term in
\eqn{cb27}. All other contributions, contained in ${\cal L}(p_0')$,
${\cal C}(p_0',q_0)$ and ${\cal Q} (p_0',q_0)$, are in fact much smaller
and thus they can be safely neglected, since they are suppressed by a
factors of the order $\Delta/\Lambda_{QCD}$. Actually the $\alpha^2$
contributions coming from radiative and Coulomb effects are even larger, or
of the same order of magnitude, than these terms, so they should be
included for consistency if ${\cal L}(p_0')$, ${\cal C}(p_0',q_0)$ and
${\cal Q} (p_0',q_0)$ are taken into account \cite{Wilk}. For the level of
accuracy of our analysis it will be sufficient to include ${\cal F}(p_0')$
and ${\cal G}(p_0',q_0)$ at order $\alpha$ only.

Evaluating numerically the integral over the phase space we then obtain for
the neutron lifetime the value \footnote{For completeness we have also
included corrections coming from the finite nucleon mass, which are
considered in section 4. However, these corrections are very small.}
$\tau^{th}_n=893.9~ sec$, which is now compatible in about 4-$\sigma$ with
the experimental value. The theoretical prediction for neutron lifetime can
be further refined by considering, for example, higher order terms in
$\alpha$, like subleading corrections, or other small effects, such as
residual average proton polarization due to parity non-conservation and so
on. Some of these effects are briefly discussed in \cite{Wilk}, but for our
purposes we do not consider them, since their small contributions would be
mainly energy independent and then account only for an overall factor. In
fact, assuming that such residual improving would eventually yield a
theoretical prediction for neutron lifetime fully compatible with the
experimental value, it is meaningful, as far as the aim of the present
investigation is concerned, to rescale the differential rate of neutron
$\beta$-decay in
\eqn{cb30} for the residual constant small factor $\tau_n^{th}/\tau_n^{exp}$.
This is a standard procedure, which allows to overcome the problem of a
precise determination of the coupling constant for weak interactions
involved in BBN, expressing this overall factor in terms of the
experimental value of neutron lifetime. We stress, however, that it is
worth rescaling the rate only after all known corrections have been
included, since this increases the accuracy of the prediction. In this way
the ansatz that the residual correction be an overall factor and not, as it
is reasonable to expect, a function of leptons energies, may introduce
errors less than $1 \%$.

The analysis of radiative corrections for $\beta$-decay shows that their
contribution to the rates for the processes in
\reac$\!\!$, relevant for BBN, is expected to be as large as few percent.
To reach accuracy of the order of one percent in rate evaluations it is
therefore necessary to correct Born rates for both radiative and Coulomb
effects. This last contribution, however, is only present when both
electron and proton are present in the initial or final state, since it can
be viewed as the electromagnetic rescattering of the two charged particles.
This results can be  also obtained diagrammatically, using the traditional
separation of electromagnetic corrections into order $\alpha$ (radiative)
and order $Z
\alpha$ (Coulomb) contributions, though this separation for the
elementary reactions in \reac is somehow arbitrary, since $Z=1$. In this
way it is straightforward to show that no Coulomb corrections ${\cal
F}(p_0')$ are present for the channels $e^+ + n \leftrightarrow \neb + p$
(see \cite{Dicus} and references therein).

Summarizing, zero temperature radiative corrections modify the Born
elementary rates in the following way:
\begin{itemize}
\item[i)] the radiative energy dependent factor ${\cal G}(p_0',q_0)$
reported in \eqn{cb22} is included in the differential rates for all
processes $(a)$-$(f)$;
\item[ii)] the Coulomb factor ${\cal F}(p_0')$ \eqn{cb27} is introduced
for processes $(a)$, $(b)$, $(e)$ and
$(f)$;
\item[iii)] all rates are rescaled by the factor
\begin{equation}
1+ \delta_\tau \equiv \tau_n^{th}/\tau_n^{exp}~~~,~~~~~~
\delta_\tau = 0.008~~~,
\label{deltatau}
\end{equation}
representing an energy independent constant correction which, as we have
discussed, should be included to reproduce the experimental value for
neutron lifetime at zero density.

\end{itemize}

\section{Finite nucleon mass corrections}
\setcounter{equation}0
In this section we consider the corrections to the Born rates for the $n
\leftrightarrow p$ reactions \reac induced by relaxing the assumption of infinite mass
for nucleons. This is a necessary improvement since the finite nucleon mass
corrections, proportional to $m_e/M_N$ or $T/M_N
\sim \alpha$, are of the same order of magnitude of the radiative ones.

The main effect \cite{main} of considering finite values for the nucleon
masses are to allow for new effective extra couplings, such as weak
magnetism, and to change both the evaluation of the spin summed squared
matrix element and the allowed phase space. Eq. (\ref{cb11}) is in fact
only valid for nucleons at rest.  Furthermore, the fact that all rates
should be averaged over the Maxwell-Boltzmann distribution for the initial
nucleon, induces corrections of order $T/M_N$. This {\it kinematical
correction} will be considered separately.

\subsection{Corrections to the transition amplitude and phase space}

For definiteness let us consider the process shown in Figure 1. At first
order in $1/M_N$, the transition amplitude can be written as
\begin{equation}\label{cm1}
M \, = \, \frac{G_F}{\sqrt{2}} \, \ov{u}_p(p) O_\mu u_n(q') \, \ov{u}_e(p')
\gamma^\mu (1 - \gamma_5) u_\nu(q)~~~,
\end{equation}
where the effective nucleon-nucleon weak coupling is given in general by
\cite{weak}
\begin{equation}\label{cm2}
O_\mu \; = \; \gamma_\mu (\cv - \ca \gamma_5) \; + \; i \, \frac{f_2}{M_N}
\, \sigma_{\mu \nu} \, (p-q')^\nu~~~.
\end{equation}
The form factor $f_2$ is the anomalous weak charged-current magnetic
moment, and we have neglected both scalar coupling, related to CVC
breaking, and pseudoscalar coupling of the nucleon, whose small value is
fixed by low-energy PCAC theorems \cite{PDG}, \cite{weak}. For the weak
magnetism term $f_2$, we have from CVC, $f_2=V_{ud}(\mu_p-\mu_n)/2=1.81
~V_{ud}$. In general, the form-factors $\cv$, $\ca$ and $f_2$, are all
dependent on the transferred four-momentum. However, at the energy scales
relevant for the considered processes, this dependence can be safely
neglected.

In appendix B we have reported the spin-summed squared modulus of the
transition amplitude (\ref{cm1}). The result is cast in a completely
general form in order to allow for a straightforward generalization to all
processes of
\reac$\!\!$.
Specializing this result for the process $\nu_e + n \rightarrow e^- + p$,
and including pure radiative effects analyzed in the previous section we
get the following expression for the rate
\bea
\omega_R( \ne  +  n  \rightarrow  e^-  +  p ) &=&
\frac{1 + \delta_\tau}{128 \pi^3 M_n}
\int_0^\infty d q_0 \int^{E_{sup}}_{E_{inf}}dp_0' ~\mod(p_0',q_0)
\nonumber\\
&{\times}&{\cal G}(p_0',q_0)~{\cal F}(p_0') ~F_\nu(q_0)~\left[ 1-F_e(p_0')
\right]~~~. \label{mm1}
\eea
where
\bea
E_{inf}&=&\frac{\left[(M_n +q_0)(M_n^2  - M_p^2 + m_e^2 + 2 q_0 M_n) - 2q_0
\xi\right]} {2M_n(M_n + 2 q_0)} ~~~,\label{mm3} \\
E_{sup}&=&\frac{\left[(M_n + q_0) (M_n^2  - M_p^2 + m_e^2 + 2 q_0 M_n) +
2q_0 \xi\right]}{2M_n(M_n + 2 q_0)}~~~,\label{mm4} \\
\xi&=&\frac 12 \sqrt{\left[M_n^2  - M_p^2 - m_e^2 + 2 q_0 M_n\right]^2 - 4
m_e^2 M_p^2} ~~~. \label{mm2}
\eea

Notice that the squared amplitude is evaluated in the neutron rest frame,
and that the result is not averaged over the neutron velocity distribution.
This effect will be considered in the following section. In Figure 3 we
plot the zero-temperature radiative correction $\Delta \omega_R
\equiv \omega_R-\omega_B$ normalized to $(\alpha / \pi)  \omega_B$ for the processes \reac.
Weak magnetism and finite nucleon mass effect on phase space are included.
The constant shift correction $\pi
\delta_\tau
/\alpha$ (see eq.
\eqn{deltatau}) has been subtracted in order to show the pure radiative
and Coulomb effects and finite mass corrections.

\subsection{Kinematical corrections}

Apart from mass corrections to the squared matrix element and allowed phase
space, there are indeed additional contributions to the reaction rates
which are of a purely kinematical nature, namely the ones connected to the
kinetic energy carried by the initial nucleon in the comoving frame.

Let us consider the reaction $\ne + n \rt e^- + p$ and, with the same
notation of the previous sections, its rate per initial nucleon
\begin{eqnarray}
\omega ( \ne  +  n  \rightarrow  e^-  +  p ) = \frac{1}{n_n} \, \int
\frac{d^3 \bvec{q}' }{(2 \pi)^3 2 q_0'} \, F_n(q_0') \, \left\{ \int
\frac{d^3 \bvec{p}}{(2 \pi)^3 2 p_0}  \, \frac{d^3 \bvec{q}}{(2 \pi)^3 2
q_0} \, \frac{d^3 \bvec{p}'}{(2 \pi)^3 2 p_0'}  \, \right. \nonumber \\
{\times} \left. (2 \pi)^4 \, \delta^{(4)} (q + q' - p - p' ) \, \mod
\right\} \, F_\nu (q_0) \left[ 1 - F_e(p_0') \right], \label{cm4}
\end{eqnarray}
where, as already mentioned, one can safely  approximate $1 \, - \,
F_p(p_0)
\, \simeq \, 1$ and
\begin{equation}\label{cm5}
n_n \; = \; 2 \, \int \frac{d^3 \bvec{q}'}{(2 \pi)^3} \, F_n(q_0')
\end{equation}
denotes the incident neutron number density. All quantities in (\ref{cm4})
are evaluated in the comoving frame, where the plasma is at rest, which
coincides with the initial nucleon rest frame only in the infinite nucleon
mass limit. Let us first note that the quantity in brace brackets is a
Lorentz invariant so that it can be equally evaluated in the neutron rest
frame. If we therefore denote with $\tilde{q}_0$ and $\tilde{p}_0'$
neutrino and electron energies in the neutron rest frame and consider the
change of variables in the integral \eqn{cm4} $(q_0,p_0') \rightarrow
(\tilde{q}_0, \tilde{p}_0')$, the transformed expression in brace brackets
is simply obtained by substituting $q_0 \rightarrow \tilde{q}_0$, $p_0'
\rightarrow \tilde{p}_0'$. To evaluate how electron and neutrino thermal
distribution factor $F_\nu (q_0) \left[ 1 - F_e(p_0')
\right]$ transforms under this change of variable, a
simple computation gives, up to $1/M_n^2$ terms
\bea
p_0' \; &\simeq& \; \tilde{p}_0' \; + \; \frac{\bvec{q}' {\cdot}
\bvec{\tilde{p}}'}{M_n} \; + \; \frac{{|\bvec{q}'|}^2}{2 M_n^2} \,
\tilde{p}_0'~~~,\label{cm10} \\
q_0 \; &\simeq& \; \tilde{q}_0 \; + \; \frac{\bvec{q}' {\cdot}
\bvec{\tilde{q}}}{M_n} \; + \; \frac{{|\bvec{q}'|}^2}{2 M_n^2} \,
\tilde{q}_0~~~, \label{cm10bis}
\eea
and thus one gets
\be
F_\nu(q_0) \, \left[ 1- F_e(p_0') \right] =  F_\nu(\tilde{q}_0) \,  \left[
1 - F_e(\tilde{p}_0') \right] \, + \,\delta \left[ F_\nu\, (1 - F_e)
\right]~~~, \label{cm5bis}
\ee
where
\begin{eqnarray}
\delta \left[ F_\nu\, (1 - F_e) \right] \simeq  F_\nu(\tilde{q}_0) \,
\left[ 1 - F_e(\tilde{p}_0') \right] \, \left\{ \frac{\bvec{q}' {\cdot}
\bvec{\tilde{p}}'}{M_n \, T} \, F_e(\tilde{p}_0') \; - \; \frac{\bvec{q}'
{\cdot}  \bvec{\tilde{q}}}{M_n \, T_\nu} \, \left[ 1 - F_\nu(\tilde{q}_0)
\right] \right. \nonumber ~~~~~~~~~\\
+ \frac{{|\bvec{q}'|}^2}{2 M_n^2 T} \, \tilde{p}_0' \, F_e(\tilde{p}_0') -
\frac{{|\bvec{q}'|}^2}{2 M_n^2 T_\nu} \, \tilde{q}_0 \, \left[
1-F_\nu(\tilde{q}_0) \right] - \frac{1}{2} \, \left( \frac{\bvec{q}'
{\cdot}  \bvec{\tilde{p}}'}{M_n \, T} \right)^2 \, F_e(\tilde{p}_0') \,
\left[ 1-2 F_e(\tilde{p}_0') \right] \nonumber \\
\left. -\frac{\bvec{q}' {\cdot}  \bvec{\tilde{p}}' \, \bvec{q}' {\cdot}
\bvec{\tilde{q}} }{M_n^2 \, T \, T_\nu} \, F_e(\tilde{p}_0') \left[
1-F_\nu(\tilde{q}_0) \right] +  \frac{1}{2} \, \left( \frac{\bvec{q}'
{\cdot}  \bvec{\tilde{q}}}{M_n \, T_\nu} \right)^2 \, \left[ 1 -
F_\nu(\tilde{q}_0) \right] \, \left[ 1 - 2 F_\nu(\tilde{q}_0) \right]
\right\}. ~~~~~\label{cm11}
\end{eqnarray}
By substituting \eqn{cm5bis} in \eqn{cm4} it is easy to see that the first
term on the r.h.s. of \eqn{cm5bis} reproduces the results of the previous
section. Hence, the new contribution, proportional to $\delta \left[
F_\nu\, (1 - F_e) \right]$ of \eqn{cm11} is an additional $1/M_n$
correction  to the Born rate which vanishes for ${|\bvec{q}'|}/M_n
\rightarrow 0$ and is of a purely kinematical origin. It is due to
the fact that the comoving and the neutron rest frames are not coinciding,
unless neutron are taken infinitely massive, and initial neutrons have a
velocity distribution in the comoving frame, over which the reaction
differential rates should be averaged.

Using that in the nonrelativistic limit
\begin{equation}\label{cm12}
F_n(q_0') = \left[\exp\left(\frac{q_0'}{T}\right) \, + \, 1\right]^{-1} \;
\simeq \; \exp\left(-\frac{q_0'}{T}\right)\; \simeq \;
\exp\left(-\frac{M_n}{T}\right) \exp\left(- \frac{{|\bvec{q}'|}^2}{2 M_n
T}\right)~.
\end{equation}
the {\it kinematical} correction is
\begin{eqnarray}
\Delta \omega_K ( \ne  +  n  \rightarrow  e^- + p ) =
\frac{e^{-\frac{M_n}{T}}}{n_n} \, \int \frac{d^3 \bvec{q}' }{(2 \pi)^3 2
q_0'} \, e^{-\frac{{|\bvec{q}'|^2}}{2 M_n T}} \, \left\{ \int \frac{d^3
\bvec{\tilde{p}}}{(2 \pi)^3 2 \tilde{p}_0}  \, \frac{d^3
\bvec{\tilde{q}}}{(2 \pi)^3 2 \tilde{q}_0} \, \frac{d^3
\bvec{\tilde{p}}'}{(2 \pi)^3 2 \tilde{p}_0'}  \, \right. \nonumber \\
{\times} \left. (2 \pi)^4 \, \delta^{(4)} (\tilde{q} + \tilde{q}' -
\tilde{p} - \tilde{p}' ) \, \mod \right\} \, \delta \left[ F_\nu\, (1 -
F_e) \right]. ~~~~~~~~~~~~ \label{cm13}
\end{eqnarray}
Since $\delta \left[ F_\nu\, (1 - F_e) \right]$ is already a $1/M_n$
correction, one can compute the above integral assuming for the quantity in
brace brackets the infinite nucleon mass limit. According to this
approximation, the dependence of the integrand function on ${\bvec{q}'}$
comes through the expression
\eqn{cm11} of $\delta \left[ F_\nu\, (1 - F_e) \right]$ only.\\ Performing
in \eqn{cm13} first the integral in $d^3{\bvec{q}'}$ and observing that
\begin{equation}\label{cm14}
\int \frac{d^3 \bvec{q}'}{(2 \pi)^3 2 M_n}  \, \exp\left\{-
\frac{{|\bvec{q}'|}^2}{2 M_n T}\right\} \, {\bvec{q}'} \; = \; 0~~~,
\end{equation}
and
\be
\frac{e^{-\frac{M_n}{T}}}{n_n}\int \frac{d^3 \bvec{q}'}{(2 \pi)^3 2 M_n}
\exp\left\{- \frac{{|\bvec{q}'|}^2}{2 M_n T}\right\} q'_i \, q'_j\, \; = \;
\frac{T}{4} \, \delta_{i j }~~~, \label{cm15}
\ee
we get the result
\begin{eqnarray}
\Delta \omega_K ( \ne  +  n  \rightarrow  e^-  +  p ) =  \frac{G_F^2 (\cvq
+ 3 \caq)}{2 \pi^3} \left( \frac{T}{2 M_n} \right) \int_0^\infty
d{|\bvec{p}'| \; |\bvec{p}'|}^2 \, q_0^2~\Theta(q_0) \, F_\nu(q_0)
~\nonumber \\ {\times} \left[ 1-F_e(p_0') \right] \left[ - 3 +
\frac{q_0^2}{T_\nu^2}
\left[ 1-F_\nu(q_0) \right] \left[1-2 F_\nu(q_0) \right]
-\frac{{|\bvec{p}'|}^2}{T^2} F_e(p_0') \left[ 1-2 F_e(p_0') \right] \right.
\nonumber \\
\left.-  3 \frac{q_0}{T_\nu} \left[1-F_\nu(q_0)\right]
+   3 \frac{p_0'}{T} F_e(p_0') + \left(\frac{\caq - \cvq}{\cvq + 3
\caq}\right)
\left(\frac{2 {|\bvec{p}'|}^2 q_0}{3 T T_\nu p_0'}\right) F_e(p_0') \left[
1-F_\nu(q_0) \right] \right]. \label{kin}
\end{eqnarray}
The above expression for $\Delta \omega_K( \ne  +  n
\rightarrow  e^-  +  p )$ cannot be extended to all reactions of
\reac by simply
using the substitution rules of Table 1. This is due to the different form
of the statistical factors involved in the corresponding expressions of
\eqn{cm4}. In Appendix C we have reported the corresponding correction
terms for the other processes $(b)$-$(f)$ in \reac. Using these expressions
the contribution $\Delta \omega_K$ can be numerically evaluated. We report
the results in Figure 4.

\section{Thermal radiative corrections}
\setcounter{equation}0
In the evaluation of the $n \leftrightarrow p$ rates, in addition to the
pure {\it electromagnetic radiative corrections}, already analyzed in the
previous sections, it is also necessary to include {\it thermal radiative
corrections} \cite{Dicus,Cambier}. These are due to finite temperature
effects in the radiative processes of Figure 5, whose strength is expected
to be of the order of $\alpha (T/m_e)$, and are related to the interactions
of the in/out particles involved in the microscopic processes $(a)$-$(f)$
with the surrounding plasma of $e^+$-$e^-$ pairs, $\ne$-$\neb$ and photons.
As in previous sections, for seek of brevity we will explicitly illustrate
the computation of corrections for the particular process $\nu_e + n
\rightarrow e^- + p$. The generalization to all six processes is
straightforward using Table 1.

To compute such thermal effects it is useful to apply the standard {\it
real time formalism} for finite temperature quantum field theory
\cite{Dolan}. In this approach propagators for $\gamma$, $e^{\pm}$, $\nu_e$ and
$\neb$ appearing in Feynman graphs of Figure 5, take the following
expression, containing a finite density term as well
\bea
i \, \Delta_\gamma^{\mu \nu}(k) &=& - \left[\frac{i}{k^2} + 2 \pi~
\frac{\delta(k^2)}{e^{\beta |k_0|} -1} \right] g^{\mu \nu} = -
\left[\frac{i}{k^2} + 2 \pi~ \delta(k^2)~B(k_0) \right] g^{\mu \nu}~~~
,\label{a1} \\ i \, S_e(p')&=& \frac{i}{\slp - m_e} - 2
\pi~\delta({p'}^2 - m_e^2)~ F_e(p_0')~(\slp + m_e)~~~,\label{a2}
\\
i \, S_\nu(q) &=&  \frac{i}{\slq} - 2 \pi~ \delta(q^2)~ F_\nu(q_0)~\slq
~~~.
\label{a3}
\eea
The additional contributions\footnote{We remind that we are assuming
vanishing chemical potential for all species.} pick up {\it real} particle
via the mass-shell $\delta$-functions and are proportional to the particle
densities in the thermal bath. They take into account the role played by
the medium in the single particle propagation. Note that a similar term for
nucleons would be suppressed by a Boltzmann factor smaller than
$\exp(-100)$ in the temperature range relevant for BBN, so it will be
neglected in the following. Hereafter we will denote with {\it thermal
radiative corrections} the finite temperature effects for all processes
corresponding to the graphs in Figure 5, once the pure radiative
contributions, considered in previous sections, have been subtracted.

The diagrams of Figures 5(a), 5(b) provide corrections of order $\alpha
(T/m_e)$ via the interference terms with the corresponding tree-level
diagram. In particular, the diagram of Figure 5(a) is responsible for
electron mass shift and wave function thermal renormalization terms. The
diagram of Figure 5(b) provides the vertex correction via its interference
with the tree-level process. Finally, in order to cure the
infrared/collinear divergences present in the 5(a) and 5(b) contributions,
it is also necessary to consider the squared amplitudes corresponding to
the diagrams of Figures 5(c)-5(f), which describe photon emission and
absorption processes.

\subsection{Electron mass shift and thermal wave function renormalization}

As already stated, the diagram of Figure 5(a) is the electron self-energy
contribution to the $\nu_e + n  \rightarrow e^- + p $ weak process and
contributes to thermal radiative corrections via electron mass and wave
function renormalization. The electron propagator reads
\be
\widetilde{S}_e = \frac{1}{\slp - m_e - \Sigma}~~~, \label{a4}
\ee
where the  self-energy $\Sigma$ is given by
\be
\Sigma(p') = i 4 \pi \alpha \int \frac{d^4 k}{(2 \pi)^4} \gamma^\mu
S_e(p'-k) \gamma^\nu \Delta_{\mu \nu}(k)~~~. \label{a5}
\ee
Hence, the {\it thermal} part of $\Sigma$ is
\be
\Sigma_T(p') = - \frac{\alpha}{\pi^2}\int d^4k \left[ \frac{2 m_e -\slp +
\slk} {(p'-k)^2 - m_e^2} ~B(k_0)~ \delta(k^2) - \frac{2 m_e +\slk}
{(p'+k)^2 }~ F_e(E)~ \delta(k^2-m_e^2)\right], \label{a6}
\ee
with $E=\sqrt{{|\bvec{k}|}^2+m_e^2}$.

By choosing the direction of electron propagation along the $3$-axis, the
thermal self-energy $\Sigma_T$ takes the simple general form
\be
\Sigma_T = \Sigma_0 + \gamma^0~ \Sigma_1 + \gamma^3~ \Sigma_2~~~,
\label{a7}
\ee
where the explicit expressions of the $\Sigma_i$ can be deduced by
comparing
\eqn{a6} with \eqn{a7}.

Differently from the usual self-energy contribution, the finite temperature
contribution contained in $\Sigma_T$ is not Lorentz invariant. This is due
to the presence of the particle densities $B(k_0)$ and $F_e(E)$, which take
the explicit expressions reported in \eqn{a6} only in the {\it comoving
frame}. Actually, the lack of Lorentz covariance, due to the presence of
the preferred frame of the medium, makes tricky the definition of a
particle state, which is generally defined as an irreducible representation
of the Poincar\`{e} group. All these difficulties originated a long debate in
literature \cite{Dicus, Cambier}, \cite{Donoghue85}-\cite{EMMP}, concerning
the correct way to evaluate the thermal wave-function renormalization,
which is strongly affected by this ambiguity. Recently, a simple approach
\cite{EMMP} has been developed in which the definition of particle
naturally comes out as the energy pole of the perturbed propagator. This
approach leads to the following expression for the renormalized projector
on positive energy states
\be
\Lambda^+_R = \left( 1 + \wsigp_1 + \frac{{|\bvec{p}'|}}{p_0'} \wsigp_2
+\frac{m_e}{p_0'} \wsigp_0 \right) \frac{\left(\slp + m^R_e \right)}{ 2
{p_0'}^R} - \frac{1}{2
p_0'}\left(\frac{\wsig_2}{{|\bvec{p}'|}}+\frac{\wsig_1}{p_0'} \right)
\left[ {\bvec{p}'} {\cdot} {\bvec{\gamma}} + \frac{{|\bvec{p}'|}^2}{
m_e}\right]~~~. \label{a7bis}
\ee
In the previous expression $p_0'=\sqrt{{|\bvec{p}'|}^2 + m_e^2}$,
${p_0'}^R=\sqrt{{|\bvec{p}'|}^2 + (m_e^R)^2}$. These notations will be
implicitly used hereafter. The {\it hat} on $\Sigma_i$ and $\Sigma_i'\equiv
d \Sigma_i/d p_0'$ reminds that they are evaluated on shell, i.e. for
$p_0'=\sqrt{{|\bvec{p}'|}^2 + m_e^2}$. Furthermore, the electron
renormalized mass $m^R_e$ results to be
\be
m^R_e = m_e + \wsig_0 + \frac{p_0'}{m_e} \wsig_1 +
\frac{{|\bvec{p}'|}}{m_e} \wsig_2 ~~~. \label{a8}
\ee
A similar result is obtained for projector on negative energy states
$\Lambda^-_R$ \cite{EMMP}.

 As it is clear from \eqn{a7bis}, at finite temperature the
projector operator does not renormalize multiplicatively. The presence of
an additional term in the expression of $\Lambda^{\pm}_R$ was already
pointed out in literature \cite{Sawyer}, but the form of this term did not
have the correct limit in the $T = 0$ covariant case. This is instead the
case of \eqn{a7bis}, as it can be simply verified. However, the additional
term in
\eqn{a7bis} does not affect the
electron wave-function renormalization contribution to rates for processes
\reac in the nonrelativistic limit for nucleons, as can be seen by explicit
calculation\footnote{To evaluate thermal radiative corrections one can
safely assume the approximation of infinite mass for nucleons.}.

For brevity we only report the expressions\footnote{We note that our result
for $\wsig_2$ differs for a numerical factor from the analogous one of Ref.
\cite{Cambier}.} for $\wsig_i$ involved in
\eqn{a8}
\bea
\wsig_0 &=& \frac{\alpha m_e}{\pi {|\bvec{p}'|}} \int_0^\infty
d{|\bvec{k}|}~\frac{{|\bvec{k}|} F_e(E)}{E} \left( \log A -\log
B\right)~~~,\label{a9} \\
\wsig_1 &=& \frac{\pi \alpha T^2}{6 {|\bvec{p}'|}}
\log\left(\frac{p_0'+{|\bvec{p}'|}}{p_0'-{|\bvec{p}'|}} \right) +
\frac{\alpha }{2 \pi {|\bvec{p}'|}}\int_0^\infty d{|\bvec{k}|}~{|\bvec{k}|}
F_e(E) \left( \log A +\log B\right)~~~,\label{a10} \\
\wsig_2 &=& \frac{\pi \alpha T^2}{3 {|\bvec{p}'|}} \left[1-\frac{p_0'} {2
{|\bvec{p}'|}} \log\left(\frac{p_0'+{|\bvec{p}'|}}{p_0'-{|\bvec{p}'|}}
\right)\right] + \frac{2 \alpha }{\pi {|\bvec{p}'|}}\int_0^\infty
d{|\bvec{k}|}~ \frac{{|\bvec{k}|}^2 F_e(E)}{E} \nonumber \\
&{\times}& \left[1 - \frac{p_0' E + m_e^2}{4 {|\bvec{k}|} {|\bvec{p}'|}}
\log A - \frac{p_0' E - m_e^2}{4 {|\bvec{k}|} {|\bvec{p}'|}} \log
B\right]~~~, \label{a11}
\eea
with
\be
A = \frac{p_0' E + m_e^2 + {|\bvec{k}|} {|\bvec{p}'|}}{p_0' E + m_e^2 -
{|\bvec{k}|} {|\bvec{p}'|}}~~~,~~~~~~~~~B = \frac{p_0' E- m_e^2 +
{|\bvec{k}|} {|\bvec{p}'|}}{p_0' E - m_e^2 - {|\bvec{k}|}
{|\bvec{p}'|}}~~~.\label{a12}
\ee
For the process $\nu_e + n \rightarrow e^- + p$ the Born graph yields the
expression for the rate for nucleon given by \eqn{cb15}. The additional
contribution due to the electron mass renormalization $\Delta \omega_M$ can
be obtained by replacing  $m_e
\rightarrow m^R_e$ in \eqn{cb15} ; thus at order
$\alpha$ we get
\bea
\Delta \omega_M (\nu_e + n \rightarrow e^- + p)= \frac{G_F^2 (\cvq + 3
\caq)}{2 \pi^3}\frac{\alpha} {\pi}\int_0^\infty
d{{|\bvec{p}'|}~{|\bvec{p}'|}}^2~ q_0^2~\Theta(q_0)~F_\nu(q_0)~
\left[1-F_e(p_0') \right] \nonumber \\
{\times} \left(\frac{2}{q_0} + \frac{F_e(p_0')}{T} -
\frac{1-F_\nu(q_0)}{T_\nu}\right)\left\{\frac{\pi^2 T^2}{3 p_0'} +
\frac{2}{p_0'}\int_0^\infty d{|\bvec{k}|}~{|\bvec{k}|}^2 \frac{F_e(E)}{E}
\right.~~~~~~~~~~ \nonumber \\
+ \left.\frac{m^2_e}{2 {|\bvec{p}'|} p_0'}\int_0^\infty
d{|\bvec{k}|}~{|\bvec{k}|}~  \frac{F_e(E)}{E}  \left(\log A -\log B\right)
\right\}~~~.~~~~~~~~~~~~~~~~~~~~~~~\label{a14}
\eea
Note that the extension of the above result to all reactions of \reac
cannot be simply done using the substitution rule of Table 1, since the
dependence on the renormalized mass comes through the statistical factors
of the Born expression as well, which depend on the particular reaction. In
Appendix C we have reported how \eqn{a14} must be modified to be applied to
the other processes $(b)$-$(f)$.

The contribution due to the electron wave function renormalization is
obtained by using the projector
\eqn{a7bis} in the evaluation of the Born rate. This procedure, having
subtracted the contribution due to the mass renormalization gives, at order
$\alpha$
\bea
\Delta \omega_W (\nu_e + n \rightarrow e^- + p)= -\frac{G_F^2 (\cvq + 3
\caq)}{2 \pi^3}\frac{\alpha} {\pi}\int_0^\infty
d{{|\bvec{p}'|}~{|\bvec{p}'|}}^2~q_0^2~\Theta(q_0)~F_\nu(q_0)~(1-F_e(p_0'))
\nonumber \\
{\times} \left\{\frac{\pi^2 T^2}{6 {|\bvec{p}'|} p_0'} \log\left(
\frac{p_0'+{|\bvec{p}'|}}{p_0'-{|\bvec{p}'|}}\right) +2 \int_0^\infty
d{|\bvec{k}|}~ {|\bvec{k}|}^2~\frac{F_e(E)}{E}
\frac{1}{{|\bvec{p}'|}^2-{|\bvec{k}|}^2} \right.~~~~~~~~~~~~~ \nonumber \\
+ \left. \frac{1}{2 {|\bvec{p}'|} p_0'} \int_0^\infty
d{|\bvec{k}|}~{|\bvec{k}|}~ \frac{F_e(E)}{E} \left[(E+p_0')\log A+ (E-p_0')
\log B \right]\right\}~~~.~~~~~~~~~~ \label{a15}
\eea
In the expression \eqn{a15} we have also subtracted an infrared divergent
term which is cancelled by an opposite term provided by the photon
absorption and emission rate. The collinear divergence, still present in
\eqn{a15} for ${|\bvec{p}'|}={|\bvec{k}|}$, together with a similar
contribution coming from the vertex correction is also compensated by
bremsstr\"{a}hlung diagrams.

Differently from \eqn{a14}, the extension of \eqn{a15} to other processes
can be straightforwardly obtained by using Table 1. The results for $\Delta
\omega_M$ and $\Delta\omega_W$, normalized to $(\alpha / \pi) \omega_B$
versus photon temperature are reported in Figures 6 and 7, respectively.

\subsection{Vertex Corrections}

The vertex correction is provided, at order $\alpha$, by the interference
term between the diagram of Figure 5(b) and the tree amplitude of Figure 1,
leading to\footnote{The result for $\Delta \omega_V$ quoted in
\cite{Cambier} as Eq. (11) has a missing factor $1/E$.}
\bea
\Delta \omega_V (\nu_e + n \rightarrow e^- + p)= \frac{G_F^2 (\cvq + 3
\caq)}{2 \pi^3} \frac{\alpha}{\pi} \int_0^\infty
d{|\bvec{p}'|}~{|\bvec{p}'|} \int_0^\infty~d{|\bvec{k}|}~ {|\bvec{k}|}
~q_0^2~\Theta(q_0) \nonumber \\
{\times} F_\nu(q_0)~ \left[1-F_e(p_0') \right] ~\frac{F_e(E)}{E}
\left\{\frac{E}{p_0'+E}\log A + \frac{E}{p_0'-E}\log B -\frac{2
{|\bvec{k}|} {|\bvec{p}'|}}{{|\bvec{p}'|}^2 -
{|\bvec{k}|}^2}\right\}.\label{a16}
\eea
The expression \eqn{a16} is extended to the other processes by using Table
1. We have reported the results in Figure 8.

\subsection{Photon Emission and Absorption}

In order to eliminate the infrared divergences due to the radiative
diagrams of Figures 5(a) and 5(b), it is necessary to include the rates of
processes in which a photon is either absorbed or emitted. In fact, they
exactly provide the divergent contributions which cancel the corresponding
ones due to vertex and wave function renormalizations. The processes give a
finite contribution as well, which takes the following form\footnote{Our
result slightly differs from Eq. (13) in Ref. \cite{Cambier}.}
\bea
\Delta \omega_\gamma(\nu_e + n \rightarrow e^- + p) = \frac{G_F^2 (\cvq + 3
\caq)}{2 \pi^3} \frac{\alpha}{\pi} \int_0^\infty
d{|\bvec{p}'|}~\int_0^\infty d{|\bvec{k}|}~ \frac{{|\bvec{p}'|}^2}
{p_0'}~B({|\bvec{k}|}) \nonumber \\
{\times}~ \left[ 1-F_e(p_0') \right]~ \left\{ - \left[ \frac{2
p_0'}{{|\bvec{k}|}}- \frac{{p_0'}^2}{{|\bvec{k}|} {|\bvec{p}'|}}
\log\left(\frac{p_0' + {|\bvec{p}'|}}{p_0' - {|\bvec{p}'|}} \right)
\right]\left[ \widetilde{Q}^2_+ + \widetilde{Q}^2_- - 2
\widetilde{Q}^2\right] \right. \label{a17} \\
-\left. \left[2S -\frac{p_0'}{{|\bvec{p}'|}} \log\left(\frac{p_0' +
{|\bvec{p}'|}}{p_0' - {|\bvec{p}'|}} \right) \right] \left[
\widetilde{Q}^2_+ - \widetilde{Q}^2_- \right] +\frac{{|\bvec{k}|}} {2
{|\bvec{p}'|}} \log\left(\frac{p_0' + {|\bvec{p}'|}}{p_0' - {|\bvec{p}'|}}
\right) \left[\widetilde{Q}^2_+ + \widetilde{Q}^2_- \right]\right\},
\nonumber
\eea
where $\widetilde{Q}^2_{\pm} \equiv (q_0{\pm} {|\bvec{k}|})^2 F_\nu(q_0
{\pm} {|\bvec{k}|})~\Theta(q_0 {\pm} {|\bvec{k}|})$, $\widetilde{Q}^2
\equiv q_0^2 F_\nu(q_0)~\Theta(q_0)$ and $S=1$. The expression \eqn{a17} is extended
to the other processes by using Table 1 and with $S=1$ for $\ne + n
\leftrightarrow e^- + p$ and $n \leftrightarrow e^- + \neb + p$, and $S=0$
for $ e^+ + n \leftrightarrow \neb + p$. The ratio $\Delta
\omega_\gamma
/ (\alpha / \pi) \omega_B$ is plotted for all processes \reac in Figure 9.
The results agree with the ones reported in Figure 8 of Ref. \cite{Cambier}
apart from the channel $e^- + p \rightarrow \neb + n$, which is there
underestimated by a factor 10. Notice that $\Delta
\omega_\gamma$ represents the leading thermal correction to neutron decay
and the inverse reaction, in the temperature range relevant for BBN. This
is simply understood since the inclusion of the process $\gamma + n
\rightarrow e^- + \neb + p$ greatly increases the neutron decay rates,
otherwise strongly suppressed by phase space.
\section{Analysis of the results and $^4He$ abundance}
\setcounter{equation}0
In the previous sections we have reported a detailed analysis of three kind
of corrections to Born rates for $n \leftrightarrow p$ processes: zero
temperature radiative corrections, finite nucleon mass and thermal
radiative effects. With reference to Figures (3), (4) and (6)-(9) the
results for each reaction channel can be summarized as follows.
\begin{itemize}
\item[(a)] $\ne + n \rightarrow e^- + p$

For the crucial BBN temperature range, $0.1~MeV \leq T \leq 3.5 ~MeV$, the
two main corrections to the Born rate come from zero temperature radiative
and kinetic terms. The contribution $\Delta \omega_R$ is weakly depending
on $T$ and represents the dominant term, though for large temperature the
kinetic correction $\Delta
\omega_K$ starts contributing significantly. The two combined
contributions correct the Born rate for a factor $6
\div 9 \%$, whilst thermal radiative ones are
of the order of $1 \%$.
\item[(b)] $e^- + p \rightarrow \ne + n $

For this channel radiative corrections are dominant at low temperature,
while the kinetic ones give a quite relevant effect in whole BBN
temperature range and correct the Born rate for a factor varying from $1
\%$ at low $T$ up to $3 \%$ for $T = 3 \div 4 ~MeV$. The radiative
contribution is quite rapidly decreasing with temperature, reaching large
negative values. This leads to a partial cancellation between $\Delta
\omega_R$ and $\Delta \omega_K$.
Thermal corrections are dominated by bremsstr\"{a}hlung contribution $\Delta
\omega_\gamma$ and can be as large as $2 \%$ of $\omega_B$ for large
temperature.
\item[(c)] $e^+ + n \rightarrow \neb + p $

For this process the radiative corrections have a behaviour quite similar
to channel $(b)$, though they are even more rapidly decreasing with
temperature. This again leads to a partial cancellation between $\Delta
\omega_R$ and the positive monotonically increasing $\Delta
\omega_K$. Thermal corrections are again mainly
provided by photon emission/absorption and monotonically increase with
temperature up to a factor $2 \%$ of the corresponding Born rate.
\item[(d)] $\neb + p \rightarrow e^+ + n $

The radiative and kinetic corrections sum up to a factor of about $5 \%$ of
the Born rate. Actually the opposite behaviour of $\Delta
\omega_R$ and $\Delta \omega_K$ conspires to give an almost constant total
correction in the whole interesting temperature range. Thermal effects are
quite small, contributing for less than $0.5 \%$.

\item[(e)] $n \rightarrow e^- + \neb + p$

For the neutron decay the radiative corrections are practically constant
and give the leading effect for small temperature, $T \leq m_e$. For larger
$T$ the thermal photon emission/absorption processes rapidly become
dominating over all other correction terms with a relative ratio to the
Born rate as large as $10^4$. However, this large correction is weakly
contributing to total $n \rightarrow p$ rate, and so to $^4He$ abundance
prediction, since in the temperature range $T > m_e$ the scattering
processes $(a)$-$(d)$ are largely dominant over decay and inverse decay.
\item[(f)] $e^- + \neb + p \rightarrow n$

Same considerations of the direct process $(e)$ hold for inverse neutron
decay. As for the direct channel, the kinetic contribution is negative down
to temperatures of the order $T \sim 0.2 ~MeV$. The contribution of
$\left|\Delta \omega_K\right|$ is however negligible  for both processes,
smaller than $1 \%$.
\end{itemize}

All results are summarized in Figure 10, where we have shown the total
relative corrections in percent. For the processes $(a)$ and $(d)$ the
correction is almost constant over the entire considered range for $T$, and
of the order of $6 \div 10 \%$ and $5 \div 6 \%$, respectively. The
positive kinetic contribution soften the deep decreasing of the radiative
terms for channels $(b)$ and $(c)$. Finally the large effect of thermal
bremsstr\"{a}hlung for neutron decay and inverse process $(e)$ and $(f)$ is
particularly evident.

We have shown the total rates for $n \leftrightarrow p$ in Figure 11, while
the total relative correction $\Delta \omega / \omega_B \equiv (
\omega - \omega_B) /\omega_B$, in percent, are plotted in Figure 12. For future applications in the BBN
codes we have performed a fit of the numerical results for $\omega(n
\rightarrow p)$ and $\omega(p
\rightarrow n)$. The fitting expressions are the following
\bea
\omega(n \rightarrow p) & = & \frac{1}{\tau_n^{exp}} \sum_{l=0}^{8} a_l
\left( \frac{T}{m_e} \right)^l~~~, \label{fitnp} \\
\omega(p \rightarrow n) & = & \frac{1}{\tau_n^{exp}} \exp
\left( -\frac{q~ m_e}{T} \right) ~\sum_{l=1}^{10} b_l
\left( \frac{T}{m_e} \right)^l~~~, \label{fitpn}
\eea
where for $n \rightarrow p$
\bea
&a_0 = 1~~;~~a_1=1.0988~~;~~a_2=-9.8297~~;~~a_3=17.379~~; ~~ a_4=32.197~~;
 \nonumber \\
 &a_5 = 27.372~~;~~a_6=-0.20975~~;~~a_7=0.67033 {\cdot} 10^{-2}~~;~~
  a_8=-0.91503 {\cdot} 10^{-4}~~~,~~~
\label{coeffnp}
\eea
while for $p \rightarrow n$
\bea
&b_1= 22.138~~;~~b_2=-77.737~~;~~b_3=130.76~~;~~b_4=-22.010~~;~~
b_5=50.262~~;
\nonumber \\
 & b_6= -4.8032~~;~~b_7=0.53087~~;~~b_8=-0.03403~~;~~
  b_9= 0.11508 {\cdot} 10^{-2}~~; \nonumber \\
  & b_{10}= -0.15850 {\cdot} 10^{-4}~~;~~q= 2.9123~~~. \label{coeffpn}
\eea
The fit has been obtained requiring that the fitting functions differ by
less than $0.1 \%$ from the numerical values.

The whole correction $\Delta \omega$ for $n \leftrightarrow p$ turn out to
be a positive decreasing function over the whole temperature range relevant
for BBN. The main contribution at low temperature for both total rates
comes from the radiative corrections, while for $T > 2
\div 3~MeV$ kinetic contribution starts dominating.
This is particularly evident by looking at Figures 13 and 14, where $\Delta
\omega_R /\omega_B$ and $\Delta \omega_K /\omega_B$ are separately plotted.
While for radiative corrections the effect on $n \rightarrow p$ total rate
is larger than on the $p \rightarrow n$ one, $\Delta
\omega_K$ shows an opposite behaviour. The competition of these two
corrections is then responsible for the presence of the inversion point in
Figure 12 at $T
\simeq 0.15 ~MeV$. Finally, the pure thermal radiative  corrections
\begin{equation}
\Delta \omega_T \equiv \Delta \omega_M + \Delta \omega_W +\Delta \omega_V
+\Delta \omega_\gamma~~~,
\label{tottherm}
\end{equation}
are plotted in Figure 15. Their order of magnitude is sensibly smaller, but
they nevertheless may contribute for a factor $0.2 \div 0.4 \%$ at the
freeze-out temperature $T \sim 1 ~MeV$. Although some differences in the
expressions for several contributions, our result for $\Delta
\omega_T$ essentially agree with the results of Ref.s \cite{Dicus,Cambier}.

The corrections on the $n \leftrightarrow p$ rates, as well as the plasma
electron mass correction to neutrino temperature, considered in Appendix A,
may produce a sensible correction to the $^4He$ mass fraction $Y_p$, which
is strongly dependent on the neutron fraction $X_n = n_n/(n_n+n_p)$ at the
nucleon freeze-out. An accurate theoretical prediction for Helium
abundance, as well as for the other light nuclei produced during BBN, can
be obtained by using the standard BBN code \cite{code} suitably modified to
take into account all the above corrections \cite{EMMP2}. Here we report
the results for the expected variation of the surviving neutron fraction
$X_n$ induced by the whole effects $\Delta \omega(n \leftrightarrow p)$.
This allows for a simple estimate of the corresponding variation of $Y_p$.

The neutron fraction is determined by the differential equation
\be
\frac{d X_n}{d T} = \frac{d t}{d T} \left[
\omega(p \rightarrow n) (1 - X_n) -
\omega(n \rightarrow p) X_n \right]~~~.
\label{diffxn}
\ee
Writing $X_n=X_n^0 + \delta X_n$, where $\delta X_n$ is the correction
induced by $\Delta \omega$, we have at first order
\begin{eqnarray}
\frac{d X_n^0}{d T} & = & \frac{d t}{d T} \left[ \omega_B(p \rightarrow n) (1 - X_n^0) -
\omega_B(n \rightarrow p) X_n^0 \right]~~~, \nonumber \\
\frac{d}{d T} \delta X_n & = & - \frac{d t}{d T} \left[
\left( \omega_B(p \rightarrow n) + \omega_B(n \rightarrow p) \right)
\delta X_n  - \Delta
\omega(p \rightarrow n) \right. \nonumber \\
& + & \left. \left( \Delta \omega(p \rightarrow n)  + \Delta \omega(n
\rightarrow p) \right) X_n^0   \right]~~~.
\label{xndxn}
\end{eqnarray}
Notice that the zero order abundance $X_n^0$ has been defined as the one
obtained by the Born amplitudes rescaled by the {\it constant} factor
$961/886.7$, which provides at tree level the correct prediction for
neutron lifetime (see our discussion in section 3). The Born rates $
\omega_B(p \leftrightarrow n)$ in \eqn{xndxn}
are therefore rescaled by the same factor. Equations (\ref{xndxn}) have
been numerically solved using our fitting function for $\omega( n
\leftrightarrow p)$ and a similar one for $\omega_B( n \leftrightarrow p)$,
which we do not report for brevity. We found for the asymptotic abundance
$\delta X_n\simeq 0.0024$, with a relative change, in percent, $\delta
X_n/X_n^0 = 1.6 \%$. It is also interesting to evaluate the effect due to
the thermal radiative contribution. Using again \eqn{xndxn}, with now the
$\Delta
\omega_T (n \leftrightarrow p)$ in the r.h.s., we get
$\delta X_n^T = -0.0002$, $\delta X_n^T /X_n^0 = -0.15 \%$.

These results allows for a simple estimation of corrections to $^4He$ mass
fraction \cite{Bernstein}
\begin{equation}
\delta Y_p = 2 ~\delta X_n~ \exp \left(- \frac{t_{ns}}{\tau_n^{exp}}
\right)~~~,
\label{he4xn}
\end{equation}
where $t_{ns} \simeq 180~ sec$ corresponds to the onset of nucleosynthesis
and the exponential factor accounts for the depletion of relic neutrons at
freeze-out due to $\beta$-decay. Using the results for $\delta X_n$ we find
\begin{eqnarray}
\delta Y_p & \simeq  - 0.0004~~,&\frac{\delta Y_p}{Y_p} = - 0.15 \%~~,~~
\mbox{ \rm thermal
~radiative} \nonumber \\
\delta Y_p & \simeq  0.004~~~,&\frac{\delta Y_p}{Y_p} = 1.6 \%~~,~~~~~~~
\mbox{\rm total}
\label{result}
\end{eqnarray}
The thermal radiative corrections decrease the value of $Y_p$, in agreement
with the results of \cite{Dicus} and \cite{Cambier}. However, the total
correction is largely dominated by zero temperature radiative and finite
mass corrections, which give a {\it positive} contribution to $Y_p$. The
total correction, evaluated with the above method, agrees with the recent
result obtained in \cite{Lopez}. They also include the effect of finite
temperature QED corrections to the equation of state of the electromagnetic
plasma \cite{Heckler} and of residual coupling of neutrinos to the thermal
bath during the $e^+$-$e^-$ annihilation phase \cite{Dicus},
\cite{Dodturn}. However, the two combined corrections give a very small
positive contribution, of the order $\delta Y_p \simeq (2 \div 3) {\cdot}
10^{-4}$.

\section{Conclusions and Outlook}
\setcounter{equation}0
In this paper we have performed a complete analysis of the corrections to
the tree level rates \reac. These reactions are relevant to fix the neutron
density at the freeze-out, which is the first step of the primordial
synthesis of light nuclei. In particular we have considered order $\alpha$
radiative corrections, $1/M_N$ effects obtained relaxing the assumption of
infinite massive nucleons, and thermal radiative effects on the rates as
well as on the behaviour of neutrino temperature after decoupling. The
total correction $\Delta
\omega$ to the rate per nucleon is a monotonically decreasing function of
the temperature in the interval $0.1 ~MeV
\leq T \leq 3.5 ~MeV$, for both processes $n \leftrightarrow p$. It varies in the range
$(2 \div 8) \%$ of the Born rates. The main effects come from radiative and
kinematical terms $\Delta \omega_R$ and $\Delta \omega_K$, considered in
section 3 and 4, respectively. We find that $\Delta
\omega_R$ contributes for about $(-1 \div 8 )\%$ in the above range for
$T$, while $\Delta \omega_K$, which originates from the relative motion of
the initial nucleon with respect to the comoving frame, grows with
temperature up to a value of about $2.5 \%$ of $\omega_B(n \leftrightarrow
p)$. Thermal radiative corrections $\Delta \omega_T$, evaluated in the real
time formalism, rapidly increase with temperature and only contribute for
about $(0.1 \div 0.7) \%$.

We have calculated how these corrections affect the theoretical prediction
for $^4He$ mass fraction $Y_p$. We have solved the differential equation
governing the evolution of neutron fraction $X_n$ and found for the total
correction induced on $Y_p$ by $\Delta \omega$ the value $\delta Y_p
\simeq 0.004$, which corresponds to a relative effect
$\delta Y_p/Y_p \simeq 1.6 \%$. This result has been obtained using the
simple method outlined in \cite{Bernstein}. The thermal radiative effects
alone contribute to this variation for a negative factor $\delta Y_p =
-0.0004$, in agreement with the results of \cite{Dicus} and \cite{Cambier}.
Finally, the total correction to $^4He$ abundance is in substantial
agreement with a recent estimate \cite{Lopez}, though the result for the
thermal radiative small correction is of the same magnitude but with an
opposite sign. A more careful analysis of the effect of $\Delta
\omega$ on Helium and all other nuclei abundances, using a modification of
the BBN codes \cite{code}, is in progress \cite{EMMP2}.

\noindent
{\large \bf Acknowledgements}

We thank E.W. Kolb for stimulating our interest to the subject considered
in this paper and for a useful discussion.

\newpage

\appendix
\section{ Neutrino temperature versus photon temperature}
\setcounter{equation}0
In this Appendix we evaluate the ratio of neutrino temperature $T_\nu$ over
the photon temperature $T$.

When weak interactions go out of equilibrium, at $T \sim 1~MeV$, neutrinos
decouple from the photon and  $e^+$-$e^-$ thermal bath. Their temperature
then decreases as $a(t)^{-1}$, with $a(t)$ the cosmological scale factor.
Shortly after, as $T$ drops below the electron mass, electrons and
positrons annihilate and transfer their entropy to photons only. To
evaluate the ratio $T_\nu / T$ we recall the expressions for the entropy
density of a particle specie $i$, with $g_i$ internal degrees of freedom
\cite{Kolb}
\begin{equation}
s_i  =  \frac{g_i}{6 \pi^2} T^3_i ~{\cal I}(x_i) =
\frac{g_i}{6 \pi^2} T^3_i \int_0^\infty \left(y^2 + 2 y x_i \right)^{1/2}
\left( 4 y^2 + 8 y x_i  + 3 x_i^2 \right)
\left( \exp (x_i+y) {\pm} 1 \right)^{-1} ~dy~~~,
\label{aaa1}
\end{equation}
where $+$, $-$ refers to fermions and bosons, respectively, and $x_i =
m_i/T_i$. Using the fact that the entropy per comoving volume is constant,
and in particular that neutrino entropy remains unchanged after decoupling
one gets
\begin{equation}
\frac{T_\nu}{T} = \left( \frac{2 \pi^4 + 15 ~{\cal I}(x_e) }
{ 2 \pi^4 + 15 ~{\cal I}(x_e^D) } \right)^{1/3}~~~,
\label{aaa2}
\end{equation}
with
\begin{equation}
x_e  =   \frac{1}{T} \left(m_e + \frac{\alpha T^2}{m_e} \right)~~~;~~~~~
x_e^D =  \frac{1}{T_D} \left(m_e + \frac{\alpha T_D^2}{m_e} \right)~~~.
\label{aaa3}
\end{equation}
The electron neutrino decoupling temperature, evaluated in \cite{Enqvist},
is $T_D \simeq 2.3~MeV$. We have verified, that changing $T_D$ in the range
$2 \div 3 ~MeV$ the ratio $T_\nu / T$ only changes for a factor less than
$0.2 \%$. We have also included the effect of thermal electron mass on
neutrino temperature via an effective simple estimate of the form $\alpha
T^2/m_e$ \cite{Dicus}.

\section{ Finite nucleon mass corrections to weak amplitudes}
\setcounter{equation}0
In this Appendix we compute the spin-summed squared amplitudes for
processes \reac for finite nucleon masses.

Let us consider the most general four-fermion effective interaction,
responsible for the process $p_1 + p_2 \rightarrow p_3 + p_4$,
\begin{equation}\label{appa1}
M \, = \, \frac{G_F}{\sqrt{2}} \, \ov{u}(p_4) O_\mu u(p_2) \, \ov{u}(p_3)
\gamma^\mu (1 - \gamma_5) u(p_1)~~~,
\end{equation}
where the effective weak coupling $O_\mu$ is given by
\begin{equation}\label{appa2}
O_\mu \; = \; \gamma_\mu (\cv - \ca \gamma_5) \; + \; i \, \frac{f_2}{M_N}
\, \sigma_{\mu \nu} \, (p_4-p_2)^\nu~~~,
\end{equation}
with $\sigma_{\mu \nu} = (i/2) [\gamma_\mu,\gamma_\nu]$. Adopting the same
notation of section 2 we obtain
\bea
\mod \; & \simeq & 8 \, ( \ca^2 \, (m_1^2 \, m_2^2  \, + \,  2 \, m_1^2 \,
m_3^2  \, + \, m_2^2 \, m_3^2  \, + \,  2 \, m_1^2 \, m_2 \, m_4  \, + \, 2
\, m_2 \, m_3^2 \, m_4 \nonumber \\
& + & m_1^2 \, m_4^2  \, + \,  2 \, m_2^2 \, m_4^2  \, + \,  m_3^2 \, m_4^2
 \, - \, 2 \, m_1^2 \, s  \, - \,  2 \, m_2^2 \, s  -  2 \, m_3^2 \, s
 \, - \,  2 \, m_4^2 \, s  \nonumber\\\, &+& \,  2 \, s^2  \, -
\, m_1^2 \, t  \, - \, m_2^2 \, t  -  m_3^2 \, t  \, - \,  2
\, m_2 \, m_4 \, t  \, - \,  m_4^2 \, t  \, +
\,  2 \, s \, t  \, + \, t^2) \nonumber \\
& + & \cv^2 \, (m_1^2 \, m_2^2  \, + \,  2 \, m_1^2 \, m_3^2  \, + \, m_2^2
\, m_3^2  \, - \,  2 \, m_1^2 \, m_2 \, m_4  \, - \, 2 \, m_2 \, m_3^2 \,
m_4 \nonumber \\
& + & m_1^2 \, m_4^2  \, + \,  2 \, m_2^2 \, m_4^2  \, + \, m_3^2 \, m_4^2
\, - \, 2 \, m_1^2 \, s  \, - \,  2 \, m_2^2 \, s
 -  2 \, m_3^2 \, s \, - \,  2 \, m_4^2 \, s  \, \nonumber\\ &+& \,  2 \,
s^2 \, - \, m_1^2 \, t  \, - \, m_2^2 \, t  -  m_3^2 \, t
\, + \,  2
\, m_2 \, m_4 \, t  \, - \, m_4^2 \, t  \, +
\,  2 \, s \, t  \, + \,  t^2) \nonumber \\
& - & 2 \, \ca \,  \frac{f_2}{M_N} \, ( m_1^2 \, m_2^3  \, - \,  m_2^3 \,
m_3^2 \,  + \, m_1^2 \, m_2^2 \, m_4 \, - \,  m_2^2 \, m_3^2 \, m_4  \, -
\, m_1^2 \, m_2 \, m_4^2 \nonumber \\ & + & m_2 \, m_3^2 \, m_4^2  \, - \,
m_1^2 \, m_4^3  \, + \, m_3^2 \, m_4^3  \, - \,  m_1^2 \, m_2 \, t  \, - \,
m_2^3 \, t \nonumber \\ & - & m_2 \, m_3^2 \, t \, - \,  m_1^2 \, m_4 \, t
\, - \,  m_2^2 \, m_4 \, t  \, - \,  m_3^2 \, m_4 \, t  \, - \,  m_2 \,
m_4^2 \, t \nonumber \\ & - & m_4^3 \, t \, + \,  2 \, m_2 \, s \, t  \, +
\,  2 \, m_4 \, s \, t
\, + \,  m_2 \, t^2  \, + \,  m_4 \, t^2) \nonumber \\
& + & 2 \,  \cv \, \ca \, ( \, - \, m_1^2 \, m_2^2  \, + \,  m_2^2 \, m_3^2
\, + \, m_1^2 \, m_4^2 \, - \,  m_3^2 \, m_4^2  \, + \, m_1^2 \, t
\nonumber \\
& + & m_2^2 \, t  \, + \, m_3^2 \, t  \, + \,  m_4^2 \, t \, - \, 2 \, s \,
t \, - \,  t^2) \nonumber \\ & + & 2 \,  \cv \, \frac{f_2}{M_N} \, (m_1^2
\, m_2^3  \, + \,  m_1^2 \, m_2 \, m_3^2 \, - \,  m_2 \, m_3^4  \, - \,
m_1^4
\, m_4  \, - \,  m_1^2 \, m_2^2 \, m_4 \nonumber \\ & + & m_1^2 \, m_3^2 \,
m_4 \, - \, m_2 \, m_3^2 \, m_4^2  \, + \,  m_3^2
\, m_4^3  \, - \,  m_1^2 \, m_2 \, s \, + \, m_2 \, m_3^2 \, s \nonumber \\
& + & m_1^2 \, m_4 \, s  \, - \, m_3^2 \, m_4 \, s \, - \, m_1^2 \, m_2 \,
t  \, - \,  m_2^3 \, t  \, + \, m_2^2 \, m_4 \, t \nonumber \\
& - & m_3^2 \, m_4 \, t  \, + \,  m_2 \, m_4^2 \, t \, - \,  m_4^3 \, t  \,
+ \, m_2 \, t^2  \, + \,  m_4 \, t^2) ).
\label{appa3}
\eea
In the above formula we have used the following definition of the two
independent Mandelstam invariants
\begin{eqnarray}
s & \equiv & \left( p_1 \, + \, p_2 \right)^2 \; = \; \left( p_3 \, + \,
p_4
\right)^2~~~,\label{appa4} \\
t & \equiv & \left( p_1 \, - \, p_3 \right)^2 \; = \; \left( p_2 \, - \,
p_4
\right)^2~~~.\label{appa5}
\end{eqnarray}
In the formula (\ref{appa3}) we have retained only the leading terms, up to
first order in the coupling $f_2$.

For the process $\nu_e + n \rightarrow e^- + p$, the corresponding $\mod$
is obtained from \eqn{appa3} by replacing $m_1$, $m_2$, $m_3$ and $m_4$
with $0$, $M_n$, $m_e$ and $M_p$, respectively. Furthermore, from the
definition of the Mandelstam invariants and using the four-momentum
conservation we have in the neutron rest frame
\begin{eqnarray}
s & = & M_n^2 + 2 M_n q_0~~~,\label{appa6} \\ t & = & M_p^2 - M^2_n + 2 M_n
(p_0' - q_0)~~~.\label{appa7}
\end{eqnarray}
By using \eqn{appa3} and changing accordingly the above correspondences one
easily obtains the expression for $\mod$ for all other processes in
\reac.

\section{Kinematical and thermal mass shift corrections
to the rates}
\setcounter{equation}0
In this Appendix we generalize to all other processes in \reac the results
reported in \eqn{kin} and
\eqn{a14} for kinematical and thermal mass corrections to Born rates
for $\ne + n \rightarrow e^- + p$.

The kinematical corrections of section 4.2 can be obtained from (\ref{kin})
by using the substitution rule of Table 1 and replacing the factor in
square brackets as follows. First of all, note that for the reaction $\neb
+ p \rt e^+ + n$ this factor is the same as in (\ref{kin}). Differently,
for $e^- + p \, \rt \, \ne + n$ and $e^+ + n
\, \rt \, \neb + p$, one has
\begin{eqnarray}
\left[-3 -\frac{q^2_0}{T_\nu^2} F_\nu(q_0) \left[ 1-2 F_\nu(q_0) \right]
+\frac{{|\bvec{p}'|}^2}{T^2} \left[ 1-F_e(p_0') \right]
\left[1-2 F_e(p_0') \right]\right.
+  3 \frac{q_0}{T_\nu} F_\nu(q_0)  \nonumber \\ \left. -  3
\frac{p_0'}{T} \left[ 1-F_e(p_0') \right] + \left(\frac{\caq -
\cvq}{\cvq + 3 \caq}\right)
\left(\frac{2 {|\bvec{p}'|}^2 q_0}{3 T T_\nu p_0'}\right) F_\nu(q_0)
\left[ 1-F_e(p_0') \right] \right] ~~~,
\end{eqnarray}
while for $n \, \rt \,  e^- + \neb + p$
\begin{eqnarray}&&
\left[ -3 -\frac{q_0^2}{T_\nu^2} F_\nu(q_0)  \left[1-2 F_\nu(q_0)
\right]
-\frac{{|\bvec{p}'|}^2}{T^2} F_e(p_0') \left[ 1-2 F_e(p_0') \right]
+ 3 \frac{q_0}{T_\nu} F_\nu(q_0)
\right. \nonumber \\  &+& \left. 3 \frac{p_0'}{T} F_e(p_0')-
\left(\frac{\caq - \cvq}{\cvq + 3 \caq}\right)
\left(\frac{2 {|\bvec{p}'|}^2 q_0}{3 T T_\nu p_0'}\right) F_e(p_0')
F_\nu(q_0) \right] ~~~.
\end{eqnarray}
Finally for $e^- + \neb + p \, \rt \, n$
\begin{eqnarray}
&& \left[ - 3 + \frac{q_0^2}{T_\nu^2} \left[ 1-F_\nu(q_0) \right]
\left[1-2
F_\nu(q_0) \right] +\frac{{|\bvec{p}'|}^2}{T^2} \left[ 1 - F_e(p_0')
\right] \left[ 1-2 F_e(p_0') \right] \right.
 -3 \frac{q_0}{T_\nu} \left[ 1-F_\nu(q_0)\right]
 \nonumber \\ &-&\left.3 \frac{p_0'}{T} \left[ 1 - F_e(p_0')\right]
- \left(\frac{\caq - \cvq}{\cvq + 3 \caq}\right)
\left(\frac{2 {|\bvec{p}'|}^2 q_0}{3 T T_\nu p_0'}\right) \left[ 1 - F_e(p_0')
\right] \left[ 1-F_\nu(q_0) \right] \right] ~~~.
\end{eqnarray}

A similar replacement is required to generalize the result of (\ref{a14}).
In particular, in addition to the substitutions of Table 1, the factor in
(\ref{a14})
\begin{equation}
\left(\frac{2}{q_0} + \frac{F_e(p_0')}{T} -
\frac{1-F_\nu(q_0)}{T_\nu}\right) ~~~,
\end{equation}
should be substituted in the following way. As for kinematical corrections,
this factor applies to the reaction $\neb + p \rt e^+ + n$ as well. For
$e^- + p
\, \rt \,
\ne + n$ and $e^+ + n \, \rt \, \neb + p$, one has instead
\begin{equation}
\left(\frac{2}{q_0} - \frac{1-F_e(p_0')}{T} +
\frac{F_\nu(q_0)}{T_\nu}\right) ~~~,
\end{equation}
while for $n \, \rt \, e^- + \neb + p$
\begin{equation}
\left(- \, \frac{2}{q_0} + \frac{F_e(p_0')}{T} -
\frac{F_\nu(q_0)}{T_\nu}\right) ~~~,
\end{equation}
and $e^- + \neb + p \rt \, n$
\begin{equation}
\left(- \, \frac{2}{q_0} - \frac{1-F_e(p_0')}{T} +
\frac{1-F_\nu(q_0)}{T_\nu}\right) ~~~.
\end{equation}

\newpage
\begin{figure}
\begin{center}
\begin{picture}(0,0)%
\includegraphics{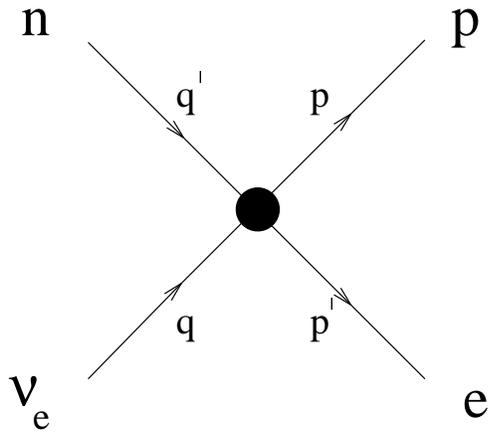}%
\end{picture}%
\setlength{\unitlength}{0.012500in}%
\begin{picture}(190,182)(50,600)
\end{picture}
\end{center}
\caption{The Feynman diagram for the reaction  $\ne + n \, \rt \, e^- +
p$ at tree level.}
\end{figure}
\begin{figure}
\epsffile{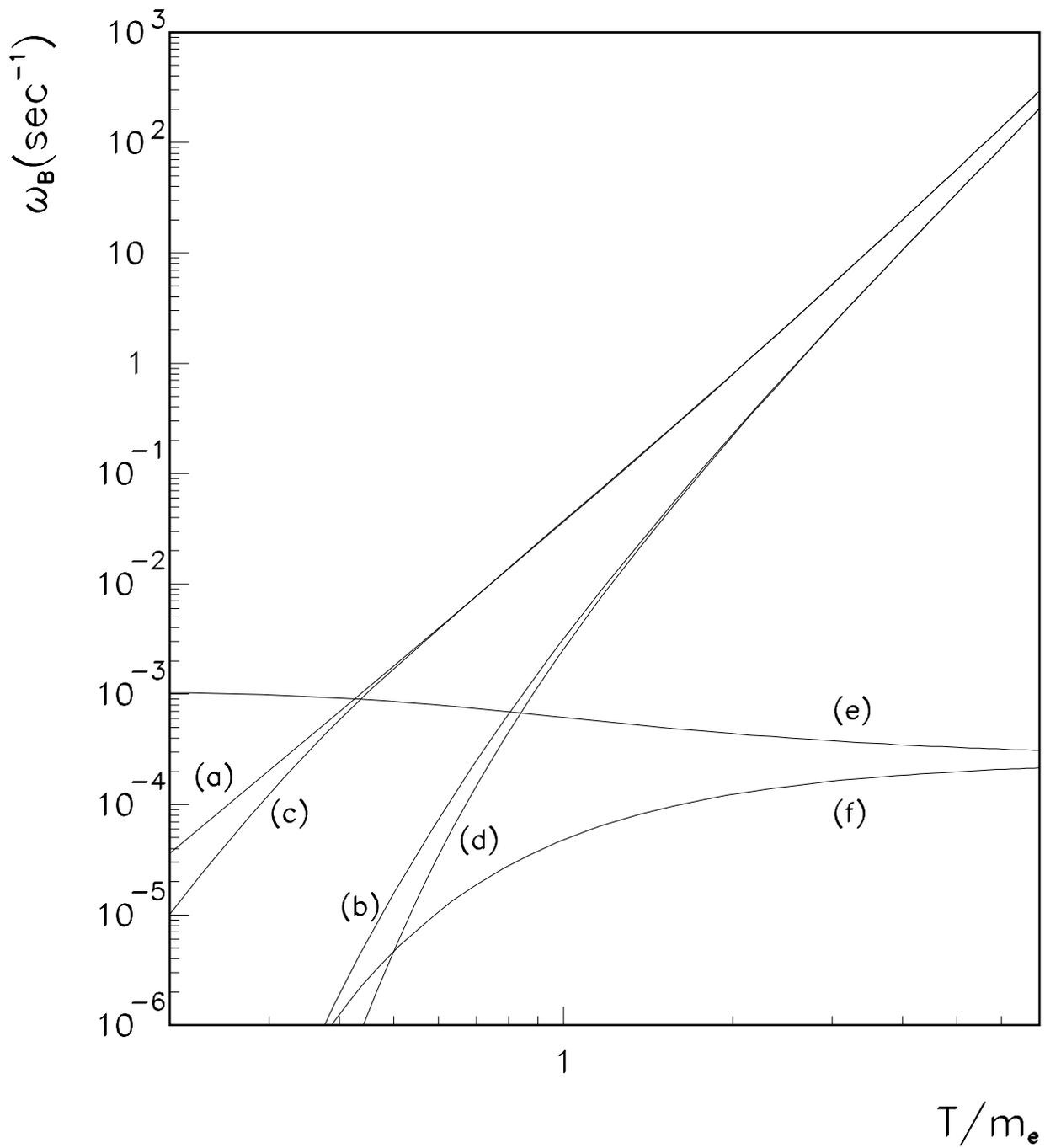}
\caption{The Born rates $\omega_B$ for all $n \leftrightarrow p$ processes
(see sect. 2). Hereafter the curves correspond to the reactions as labelled
in \reac.}
\end{figure}
\begin{figure}
\epsffile{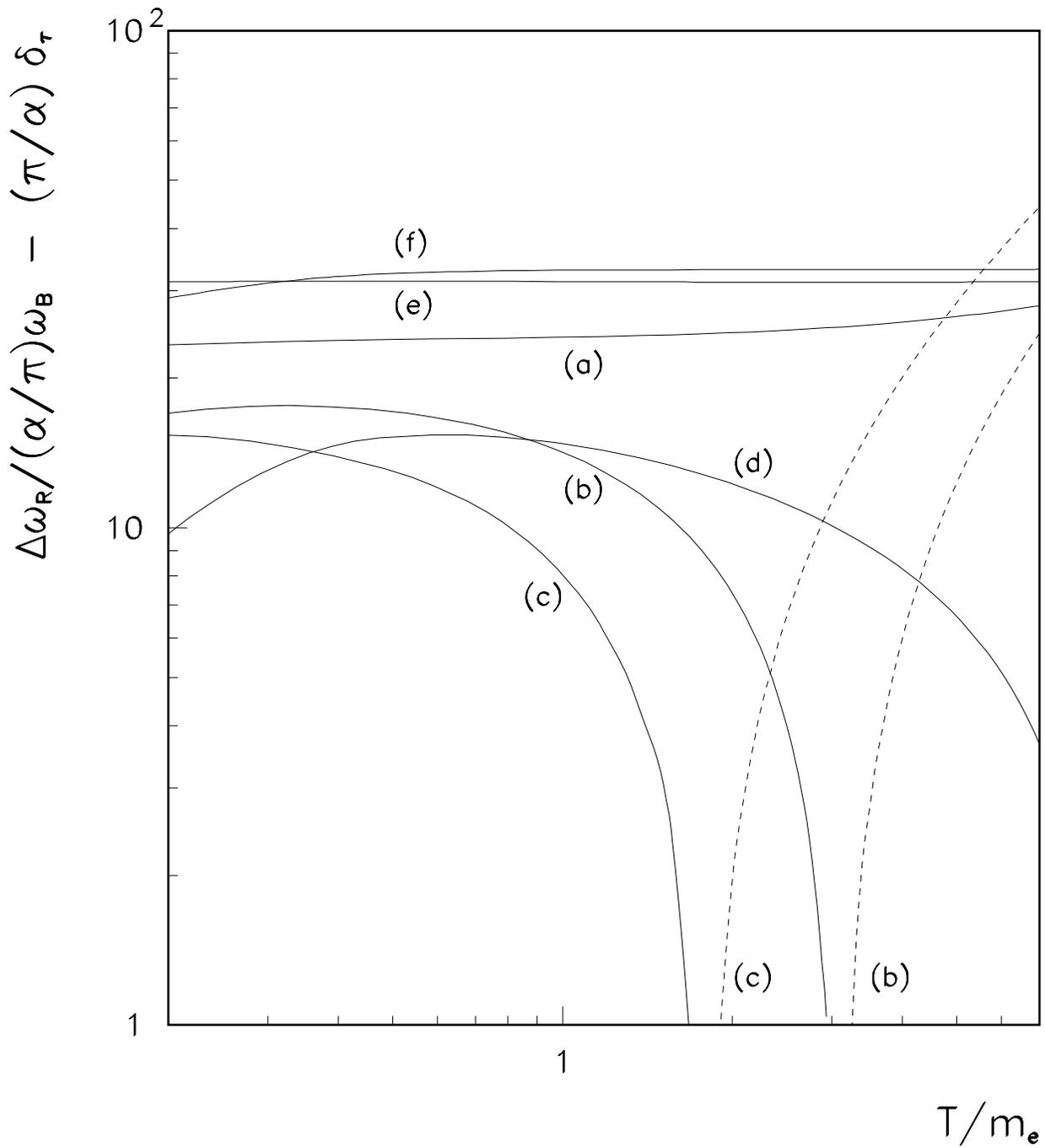}
\caption{The quantity $\Delta \omega_R /(\alpha/\pi) \omega_B -
\pi \delta_\tau / \alpha$
(see sect. 4.1) versus photon temperature. Hereafter dashed lines
correspond to negative values.}
\end{figure}
\begin{figure}
\epsffile{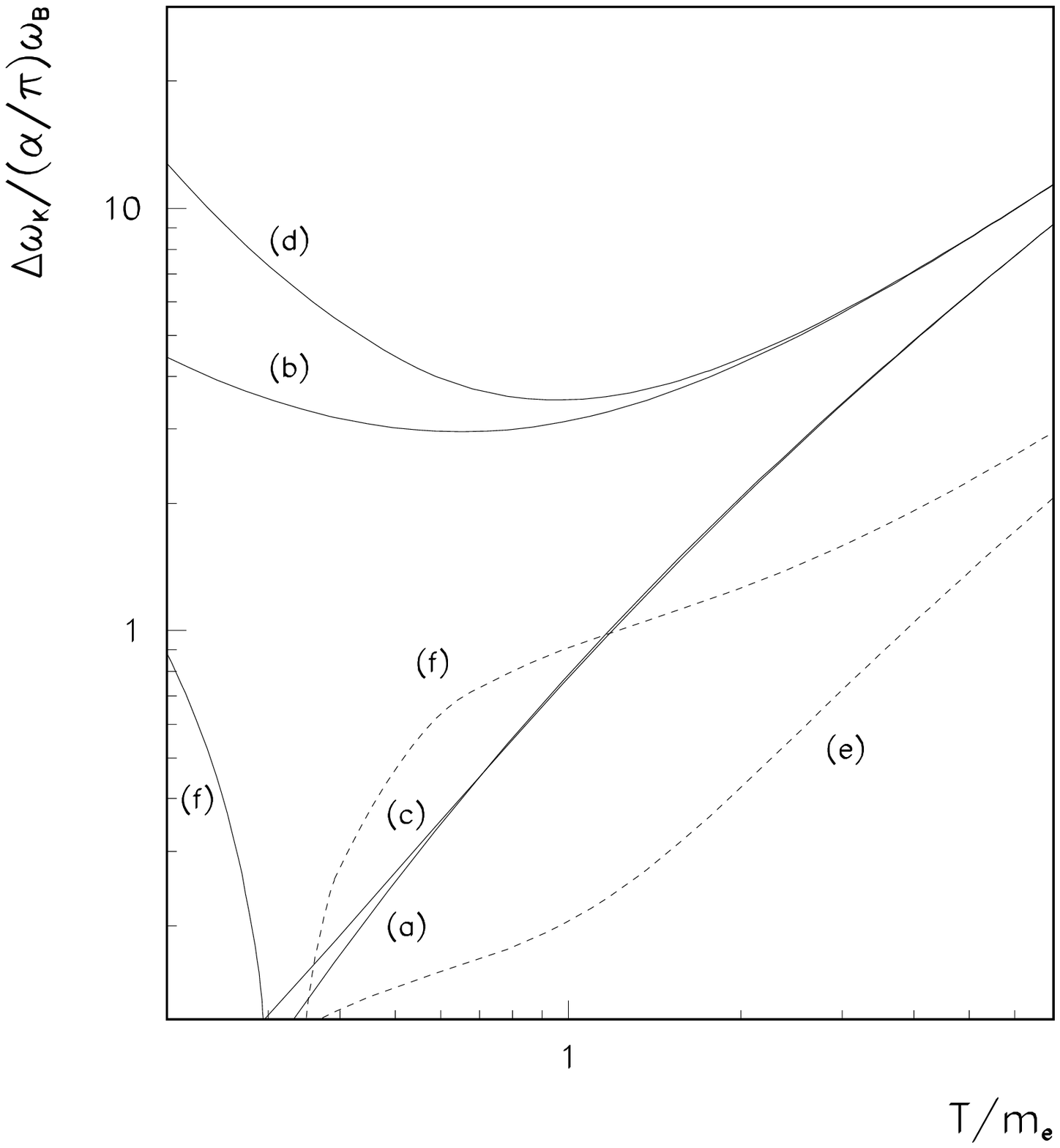}
\caption{The ratio $\Delta \omega_K/(\alpha/\pi) \omega_B$ (see sect. 4.2).}
\end{figure}
\begin{figure}
\begin{center}
\begin{picture}(0,0)%
\includegraphics{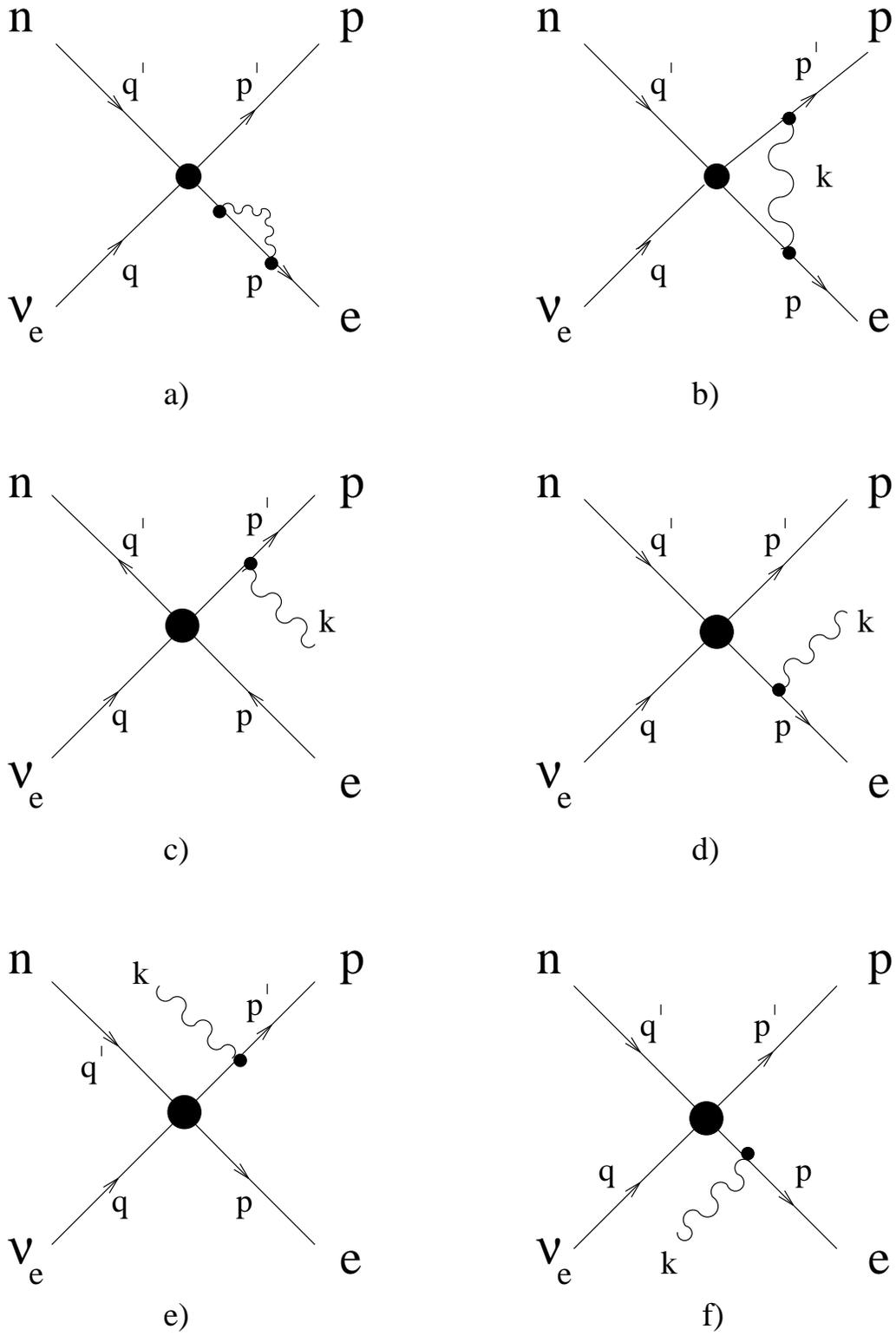}%
\end{picture}%
\setlength{\unitlength}{0.012500in}%
\begin{picture}(427,642)(65,140)
\end{picture}
\end{center}
\caption{The one-loop and photon emission/absorption diagrams for the
reaction  $\ne + n \, \rt \, e^- +p$.}
\end{figure}
\begin{figure}
\epsffile{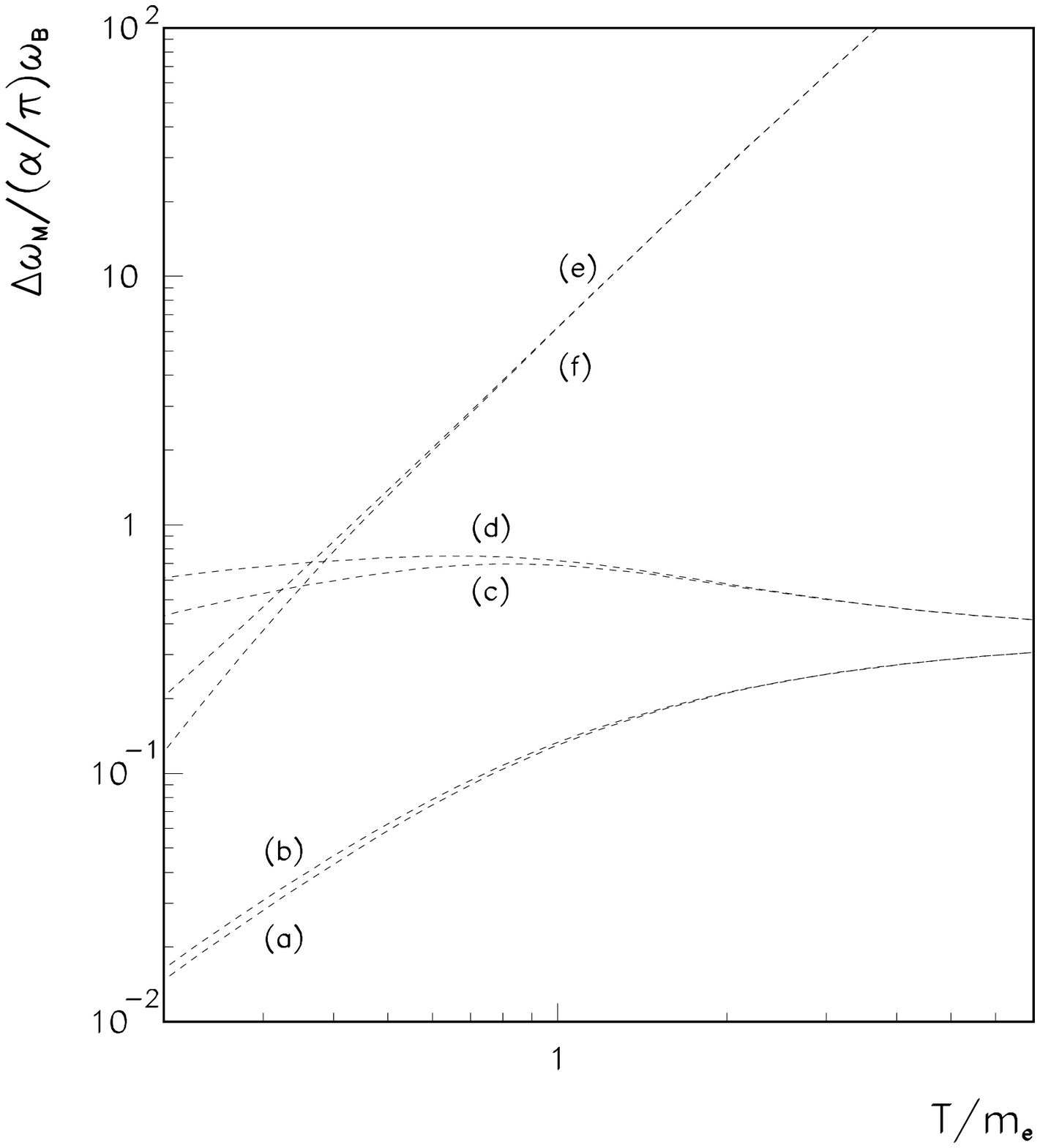}
\caption{The ratio $\Delta \omega_M/(\alpha/\pi) \omega_B$ (see sect. 5.1).}
\end{figure}
\begin{figure}
\epsffile{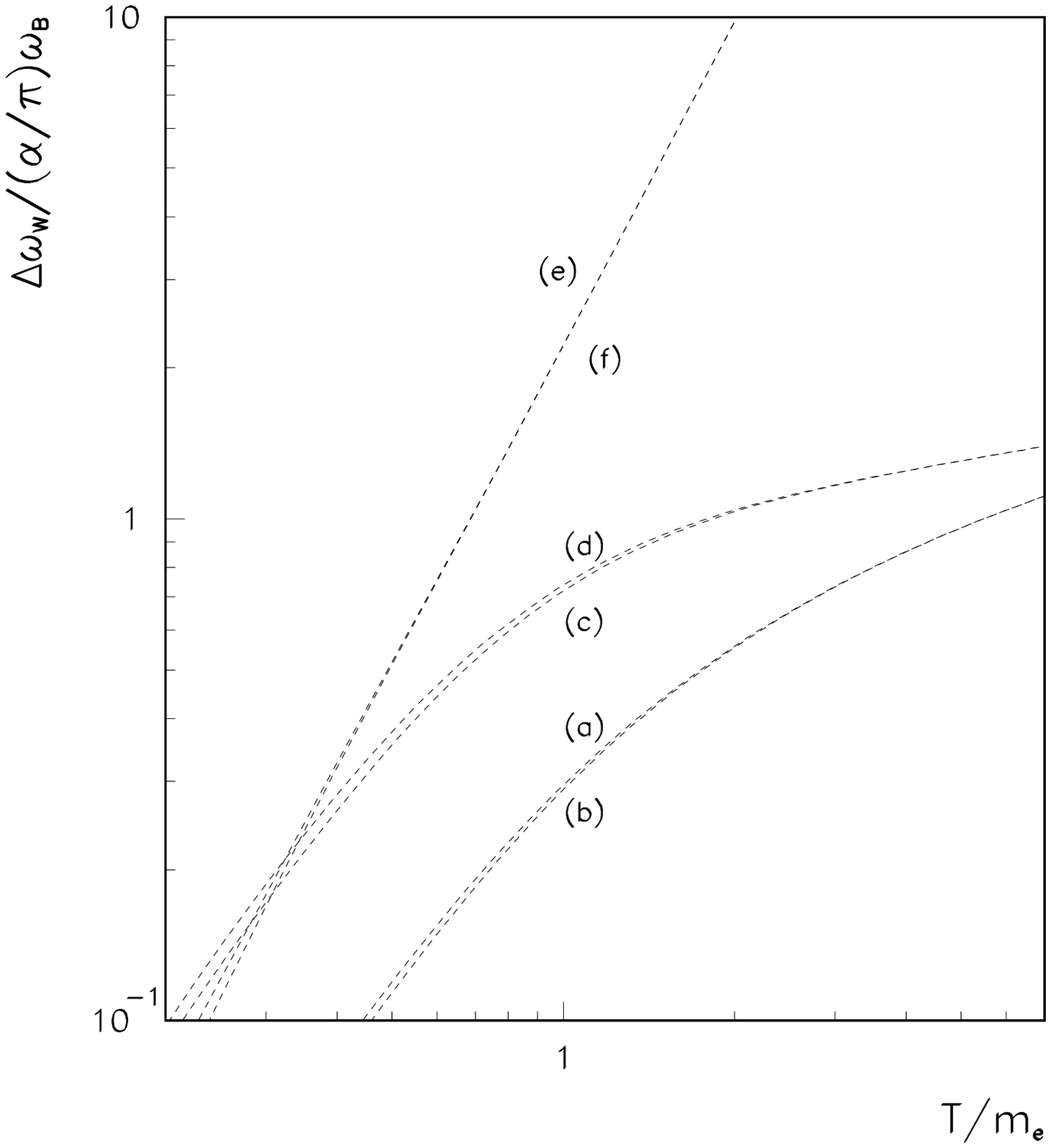}
\caption{The ratio $\Delta \omega_W/(\alpha/\pi) \omega_B$ (see sect. 5.1).}
\end{figure}
\begin{figure}
\epsffile{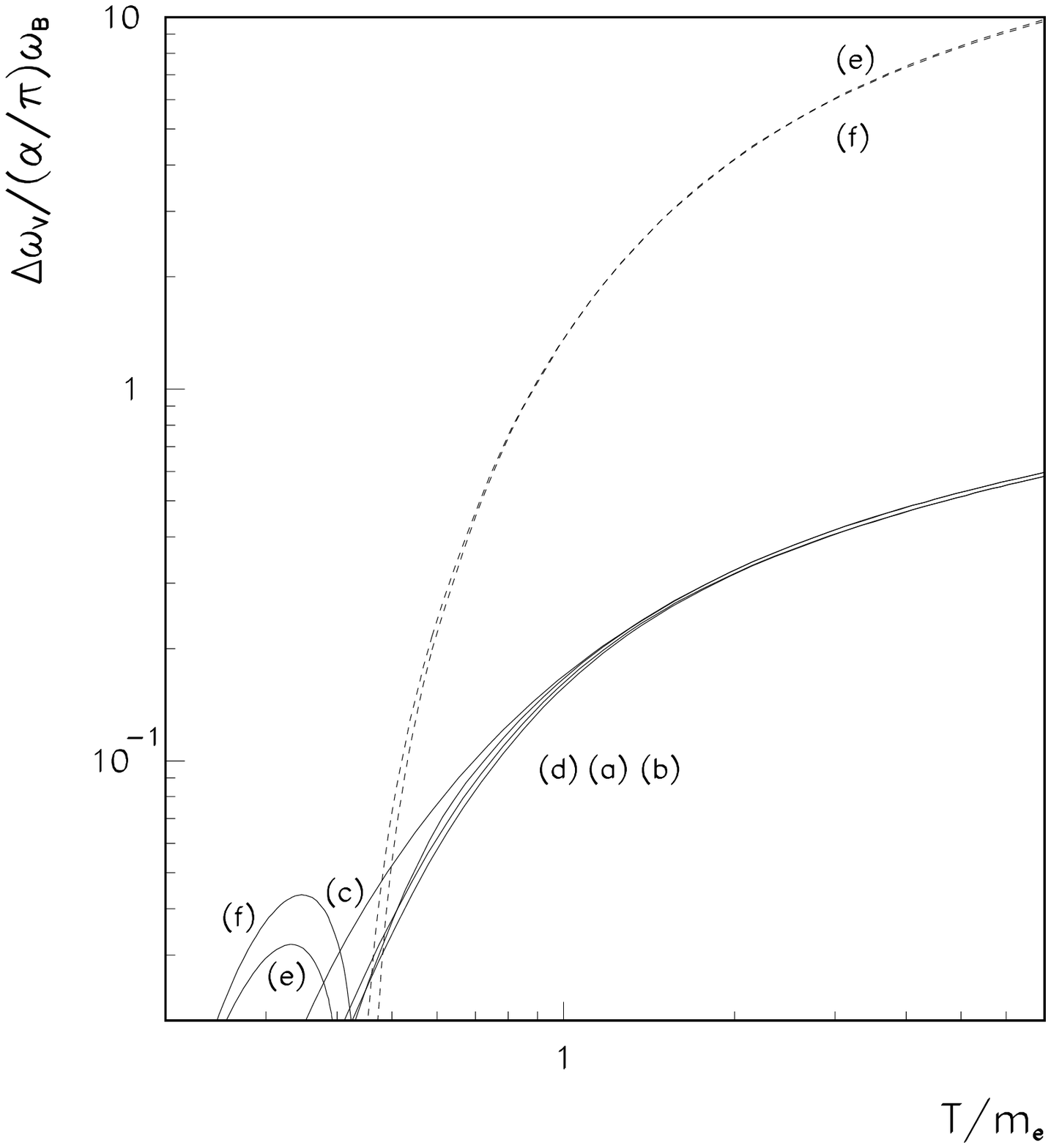}
\caption{The ratio $\Delta \omega_V/(\alpha/\pi) \omega_B$ (see sect. 5.2).}
\end{figure}
\begin{figure}
\epsffile{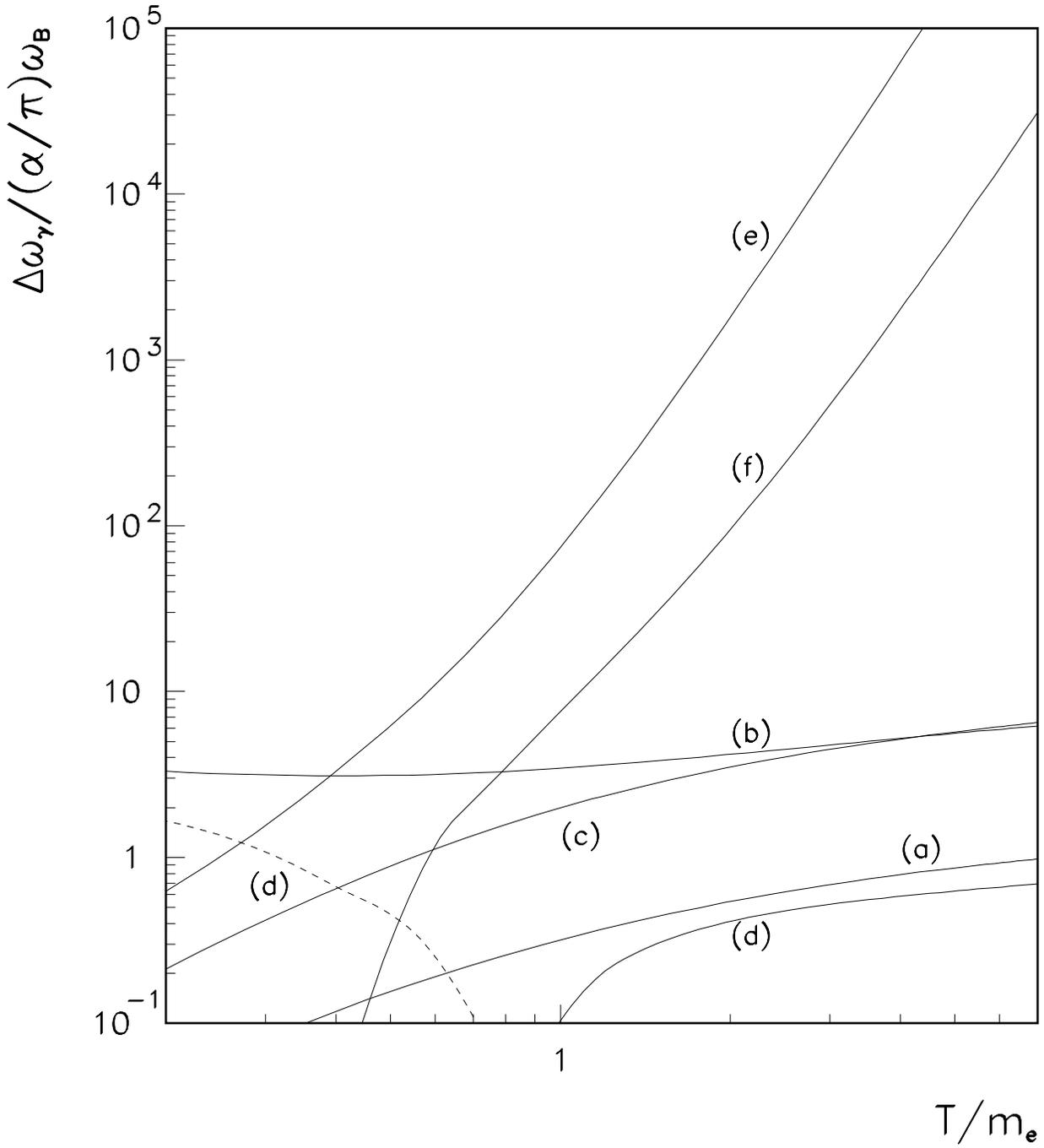}
\caption{The ratio $\Delta \omega_\gamma/(\alpha/\pi) \omega_B$
(see sect. 5.3).}
\end{figure}
\begin{figure}
\epsffile{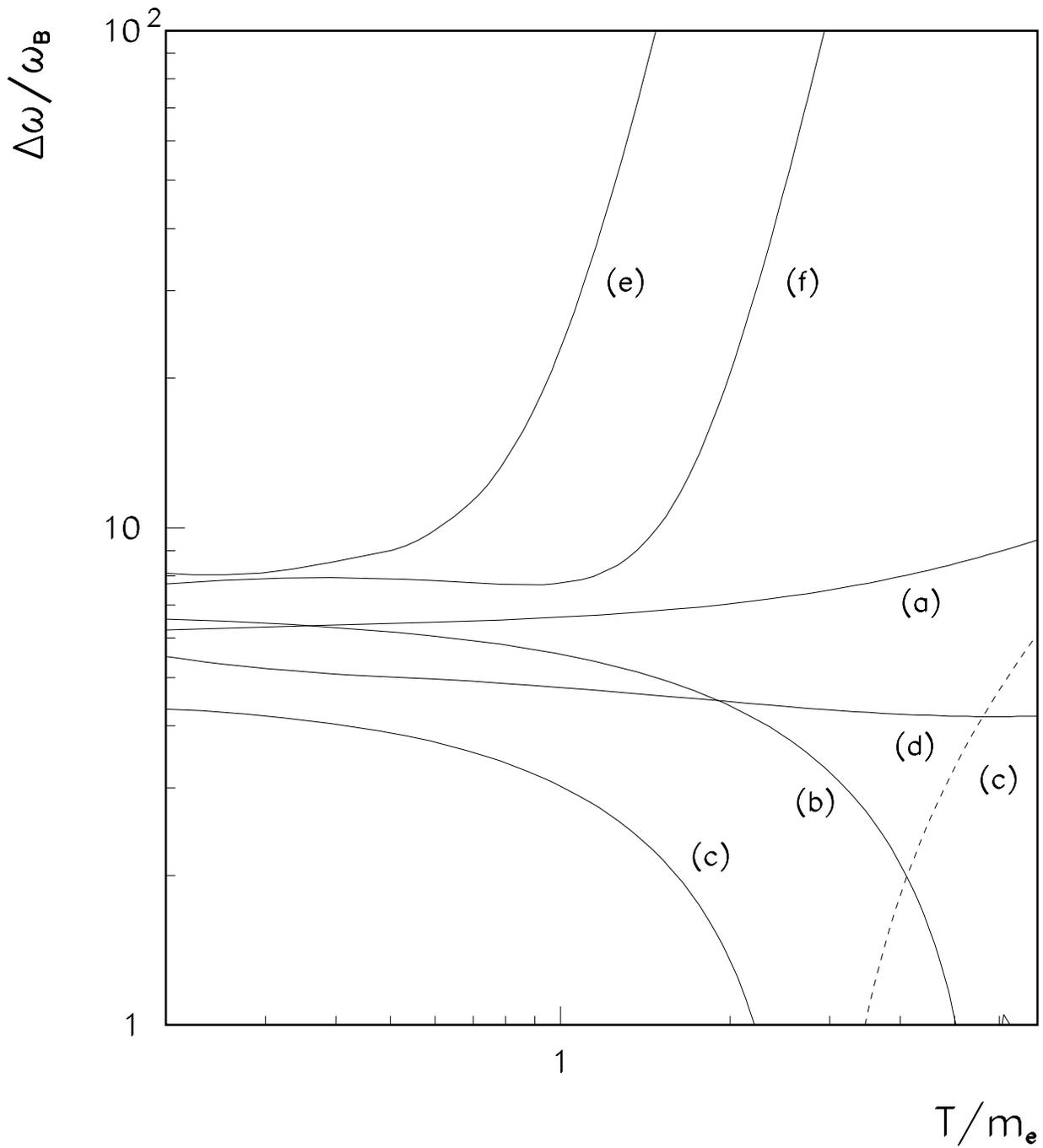}
\caption{The total relative correction to the Born rates, in percent,
are shown for the six processes $(a)$-$(f)$.}
\end{figure}
\begin{figure}
\epsffile{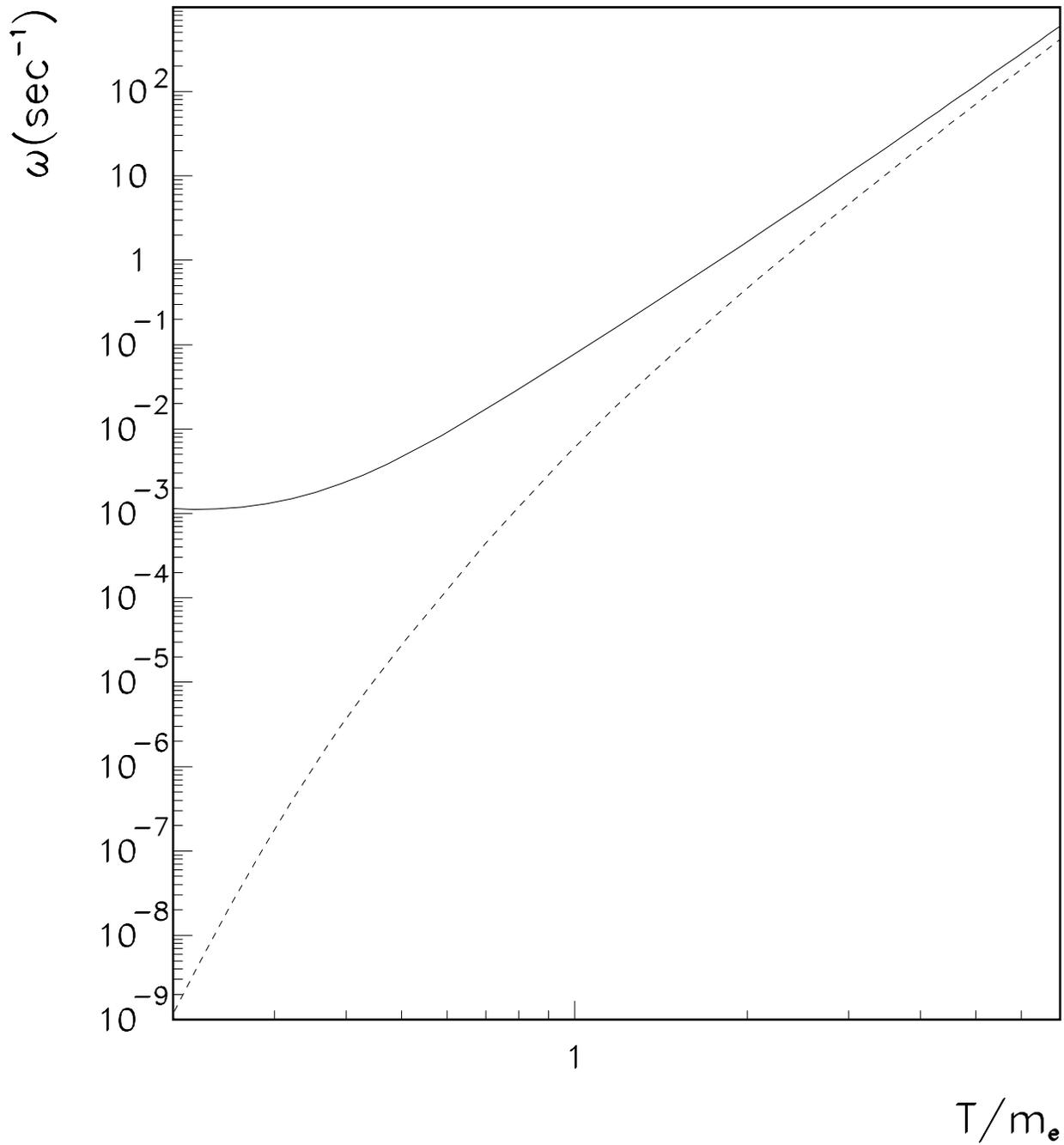}
\caption{The total rates $\omega(n \rightarrow p)$  (solid line) and
$\omega(p \rightarrow n)$  (dashed line), including all radiative, finite
mass and thermal corrections.}
\end{figure}
\begin{figure}
\epsffile{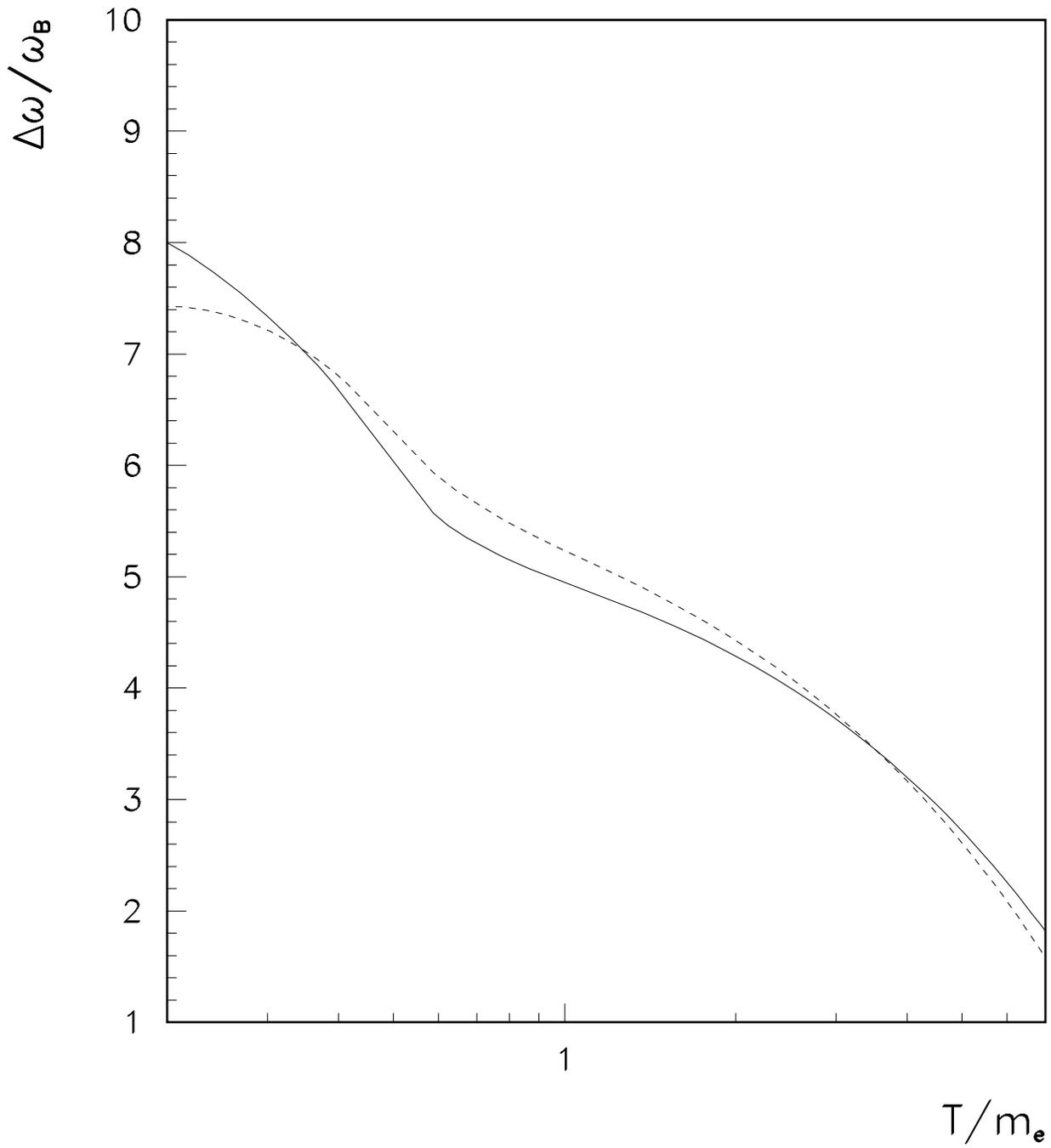}
\caption{The total relative corrections $\Delta \omega /\omega_B$ for the
$n \rightarrow p$ (solid line) and $p \rightarrow n$ (dashed line)
processes, expressed in percent.}
\end{figure}
\begin{figure}
\epsffile{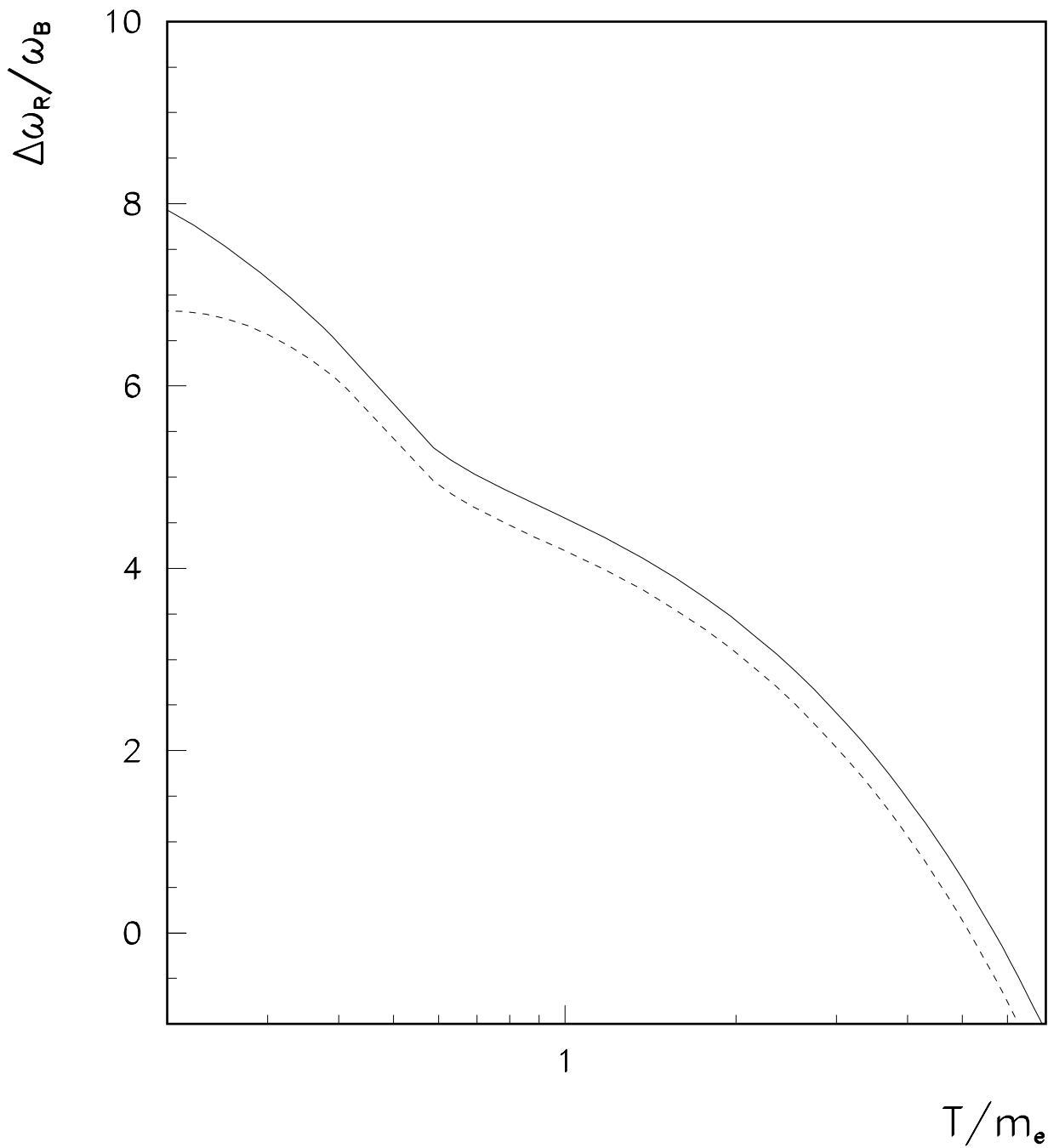}
\caption{The zero temperature radiative corrections, for $n \rightarrow p$ (solid line) and $p \rightarrow n$
(dashed line) processes. Finite mass contributions coming from phase space
integration and weak magnetism are also included. The result is expressed
in percent.}
\end{figure}
\begin{figure}
\epsffile{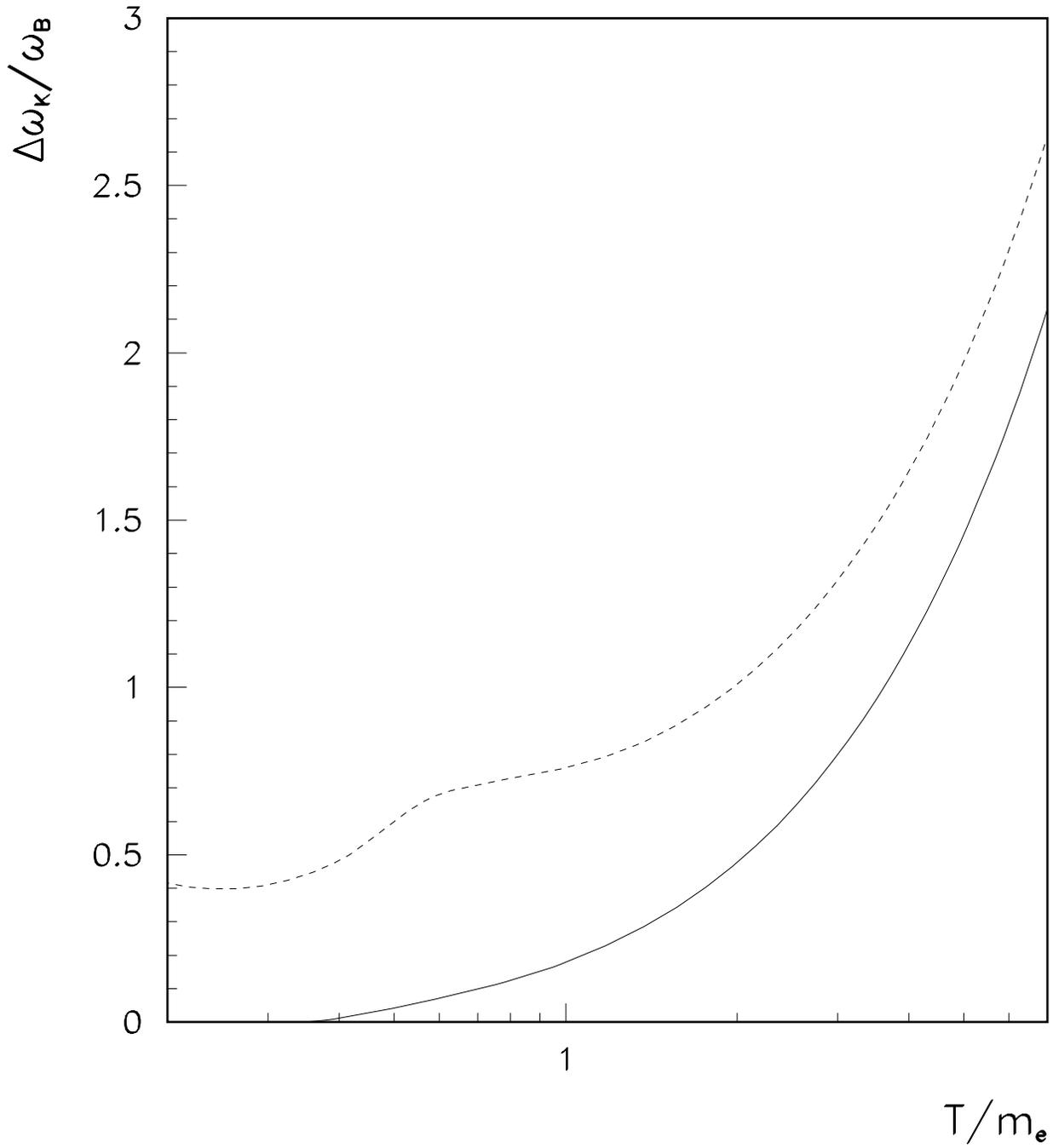}
\caption{The
kinematical contributions  to the $n \rightarrow p$ (solid line) and $p
\rightarrow n$ (dashed line) processes, normalized to the Born rates,
in percent.}
\end{figure}
\begin{figure}
\epsffile{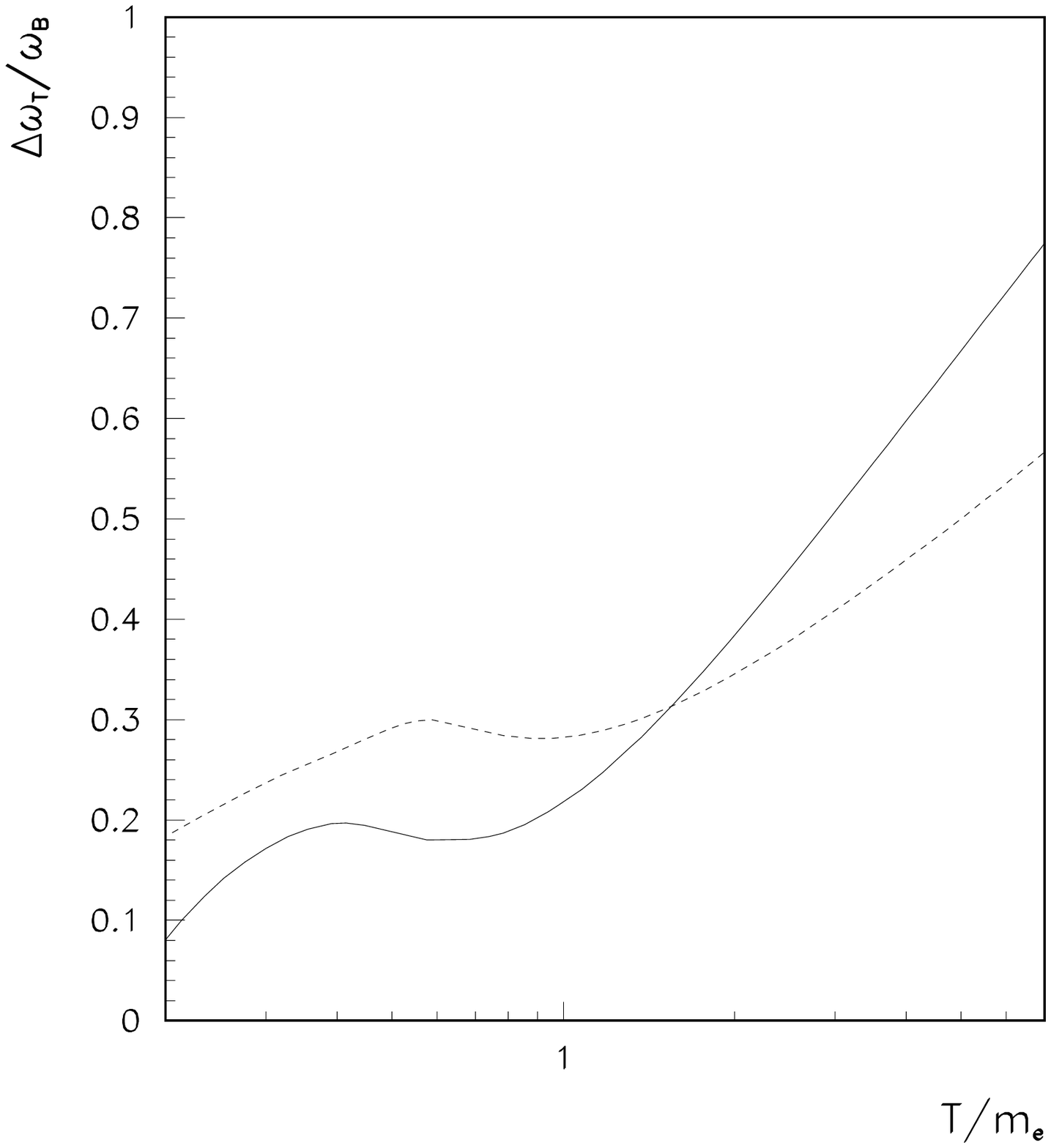}
\caption{The whole thermal radiative corrections to
the $n \rightarrow p$ (solid line) and $p
\rightarrow n$ (dashed line) processes, expressed in percent.}
\end{figure}


\begin{thebibliography} \\

\bibitem{revBBN} P.J.E. Peebles, {\em Physical Cosmology}, (Princeton
University Press, 1971, Princeton);\\
 S. Weinberg, {\em Gravitation and Cosmology}, (Wiley, 1972, New York);\\
 D.N. Schramm and R.V. Wagoner, {\it Ann. Rev. Nucl. Sci.} {\bf 27} (1977)
 37.

\bibitem{Kolb} E.W. Kolb and M.S. Turner, {\em The Early Universe},
(Addison-Wesley Publishing Company, 1990, New York).

\bibitem{BBNdata} K.A. Olive, Proc. of the Workshop TAUP97, Gran Sasso
Laboratory, Italy, september 7-11, 1997, astro-ph/9712160;\\
 G. Steigman, astro-ph/9803055.

\bibitem{HE4data} K.A. Olive and D. Thomas, {\it Astropart. Phys.} {\bf 7}
(1997) 27;\\
 Y.I. Izotov, T.X. Thuan, and V.A. Lipovetsky, {\it Ap. J. Suppl.} {\bf
 108} (1997) 1;\\
 R.F. Carswell, M. Rauch, R.J. Weymann, A.J. Cooke and J.K. Webb, {\it MNRAS} {\bf 268}
 (1994) L1;\\
 A. Songaila, L.L. Cowie, C. Hogan and M. Rugers, {\it Nature} {\bf 368} (1994) 599;\\
 S. Burles and D. Tytler, Proc. of the ISSI Workshop {\it Primordial
 Nuclei and their Galactic Evolution}, astro-ph/9712265.

\bibitem{Sarkar} S. Sarkar, {\it Rept. Prog. Phys.} {\bf 59} (1996) 1493.

\bibitem{Nuclear} L.M. Krauss and P. Romanelli, {\it Ap. J.} {\bf 358} (1990) 47;\\
P.J. Kernan and L.M. Krauss, {\it Phys. Rev. Lett.} {\bf 72} (1994) 3309;\\
G. Fiorentini, E. Lisi, S. Sarkar and F.L. Villante, astro-ph/9803177.

\bibitem{code} R.V. Wagoner, {\it Ap. J. Suppl.} {\bf 18} (1969) 247;
{\it Ap. J.} {\bf 179} (1973) 343;\\
 L. Kawano, preprint FERMILAB-Pub-88/34-A; preprint
 FERMILAB-Pub-92/04-A.

\bibitem{Sirlin} A. Sirlin, {\em Phys. Rev.} {\bf 164} (1967) 1767.

\bibitem{Marciano} W.J. Marciano and A. Sirlin, {\em Phys. Rev. Lett.}
   {\bf 56} (1986) 22, and references therein.

\bibitem{MS} W.J. Marciano and A. Sirlin, {\em Phys. Rev. Lett.}
   {\bf 46} (1981) 163.

\bibitem{Wilk} D.M. Wilkinson, {\em Nucl. Phys.} {\bf A377} (1982) 474.

\bibitem{PDG} C. Caso et al., {\it Eur. Phys. Jour.} {\bf C3} (1998) 1.

\bibitem{Dicus} D.A. Dicus, E.W. Kolb, A.M. Gleeson, E.C.G. Sudarshan,
V.L. Teplitz and M.S. Turner, {\it Phys. Rev.} {\bf D26} (1982) 2694.

\bibitem{Cambier} J.L. Cambier, J.R. Primack and M. Sher, {\it Nucl. Phys.}
{B209} (1982) 372.

\bibitem{Dolan}
L. Dolan and R. Jackiw, {\it Phys. Rev.} {\bf D9} (1974) 3320.

\bibitem{Donoghue85} J.F. Donoghue, B.R. Holstein and R.W. Robinett, {\it
Ann. Phys.} (N.Y.) {\bf 164} (1985) 23.

\bibitem{Donoghue} J.F. Donoghue and B.R. Holstein, {\it Phys. Rev.}
{\bf D28} (1983) 340; {\it Phys. Rev.} {\bf D29} (1984) 3004.

\bibitem{Johansson} A.E. Johansson, G. Peresutti and B.S. Skagerstam,
{\it Nucl. Phys.} {B278} (1986) 324.

\bibitem{Keil} W. Keil, {\it Phys. Rev.} {\bf D40} (1989)1176.

\bibitem{Baier} R. Baier, E. Pilon, B. Pire and D. Schiff, {\it Nucl.
Phys.} {\bf B336} (1990) 157.

\bibitem{KeilKobes} W. Keil and R. Kobes, {\it Physica} {\bf A158} (1989)
47.

\bibitem{LeBellac} M. LeBellac and D. Poizat, {\it Z. Phys.} {\bf C47}
(1990) 125.

\bibitem{Altherr} T. Altherr and P.Aurenche, {\it Phys. Rev.} {\bf D40} (1989)
4171.

\bibitem{Kobes} R.L. Kobes and G.W. Semeneff, {\it Nucl. Phys.} {\bf B260}
(1985) 714; ibidem {\bf B272} (1986) 329.

\bibitem{Sawyer} R.F. Sawyer, {\it Phys. Rev.} {\bf D53} (1996) 4232.

\bibitem{Chapman} I.A. Chapman, {\it Phys. Rev.} {\bf D55} (1997) 6287.

\bibitem{EMMP} S. Esposito, G. Mangano, G. Miele and O. Pisanti,
{\it Phys. Rev.} {\bf D58} (1998) 105023.

\bibitem{Dodturn} M.A. Herrera and S. Hacyan. {\it Ap. J.} {\bf 336}
(1989) 539;\\
 N.C. Rana and B. Mitra, {\it Phys. Rev.} {\bf D44} (1991) 393;\\
 S. Dodelson and M.S. Turner, {\it Phys. Rev.} {\bf D46} (1992) 3372; \\
 A.D. Dolgov and M. Fukugita, {\it Phys. Rev.} {\bf D46} (1992) 5378; \\
 N.Y. Guedin and O.Y. Guedin, astro-ph/9712199;\\
 S. Hannestad and J. Madsen, {\it Phys. Rev.} {\bf D52} (1995) 1764; \\
 A.D. Dolgov, S.H. Hansen and D.V. Semikoz, {\it Nucl. Phys.} {\bf B503}
 (1997) 426;\\
 B. Fields, S. Dodelson and M.S. Turner, {\it Phys. Rev.} {\bf D47} (1993)
 4309.

\bibitem{main}
D. Seckel, preprint BA-93-16, hep-ph/9305311;\\ R. E. Lopez, M. S. Turner
and G. Gyuk, {\it Phys. Rev.} {\bf D56} (1997) 3191.

\bibitem{Kubo} V.P. Gudkov and K. Kubodera, nucl-th/9706074.

\bibitem{weak} F. Halzen and A.D. Martin, {\em Quarks and Leptons}, (Wiley, 1984, New York).

\bibitem{EMMP2} S. Esposito, G. Mangano, G. Miele and O. Pisanti,
in preparation.

\bibitem{Bernstein} J.Bernstein, L.S. Brown and G. Feinberg, {\it Rev. Mod. Phys.} {\bf 61}
(1989) 25.

\bibitem{Lopez} R.E. Lopez and M.S. Turner,
preprint FERMILAB-Pub-98/232-A (1998), astro-ph/9807279.

\bibitem{Heckler} A.J. Heckler, {\it Phys. Rev.} {\bf D49} (1994) 611.

\bibitem{Enqvist} K. Enqvist, K. Kainulainen and V. Semikoz,
{\it Nucl. Phys.} {\bf B374} (1992) 392.
\end{thebibliography}
\end{document}